\documentclass[a4paper,11pt]{article}
\pdfoutput=1 
\usepackage{jheppub}
\usepackage{graphicx}
\usepackage{amsfonts}
\usepackage{hhline}
\usepackage{multirow}
\usepackage{array}
\usepackage{amsmath}
\usepackage{amssymb}
\usepackage{maybemath}
\usepackage{float}
\usepackage[lofdepth,lotdepth]{subfig}
\usepackage[countmax]{subfloat}
\usepackage{cancel}
\usepackage[top=5cm,bottom=1.5cm,left=4.65cm,right=2.50cm]{geometry}
\def\beq{\begin{equation}}
\def\eeq{\end{equation}}
\def\bea{\begin{eqnarray}}
\def\eea{\end{eqnarray}}

\def\roughly#1{\mathrel{\raise.3ex\hbox
{$#1$\kern-.75em\lower1ex\hbox{$\sim$}}}}

\def\tbslash{\tbar\hspace{-10pt}\not{}\hspace{4pt}}
\def\tslash{t\hspace{-10pt}\not{}\hspace{4pt}}

\def\nslash{n\hspace{-10pt}\not{}\hspace{4pt}}

\def\vmet{\vec{E}_{T}\hspace{-18pt}\not{}\hspace{12pt}}

\def\tbar{\bar{t}}
\def\bbar{\bar{b}}
\def\qbar{\bar{q}}
\def\nubar{{\bar{\nu}}_{\ell}}

\def\ppprocess{pp\to t\,\left(\rightarrow b {\ell}^+ \nu_{\ell}\right) \tbar\,\left(\rightarrow\bbar {\ell}^-\nubar\right)\,H}

\def\kp{\kappa_t}
\def\kpt{\tilde{\kappa}_t}

\def\TPa{\epsilon(t,\tbar,n_t,n_{\tbar})}
\def\TPb{\epsilon(Q,\tbar,n_t,n_{\tbar})}
\def\TPc{\epsilon(Q,t,n_t,n_{\tbar})}
\usepackage[normalem]{ulem}
\usepackage{color}
\definecolor{BrickRed}{cmyk}{0,0.89,0.94,0.28}
\definecolor{DarkGreen}{cmyk}{1,0,1,0.5}
\definecolor{Blue}{cmyk}{1,1,0,0}
\definecolor{BurntOrange}{cmyk}{0,0.51,1,0}

\def\soutdl{\bgroup\markoverwith{\textcolor{BrickRed}{\rule[0.5ex]{2pt}{0.4pt}}}\ULon}
\def\soutps{\bgroup\markoverwith{\textcolor{DarkGreen}{\rule[0.5ex]{2pt}{0.4pt}}}\ULon}
\def\soutkk{\bgroup\markoverwith{\textcolor{Blue}{\rule[0.5ex]{2pt}{0.4pt}}}\ULon}
\def\soutas{\bgroup\markoverwith{\textcolor{BurntOrange}{\rule[0.5ex]{2pt}{0.4pt}}}\ULon}

\title{{\boldmath Pseudoscalar top-Higgs coupling:  Exploration of $\mathrm{CP}$-odd observables to 
resolve the sign ambiguity
}}

\author[a]{Nicolas Mileo}
\author[b]{\hspace*{-4pt}, Ken Kiers}
\author[a]{\hspace*{-4pt}, Alejandro Szynkman}
\author[b]{\hspace*{-4pt}, Daniel Crane}
\author[b]{and Ethan Gegner}

\affiliation[a]{{\it IFLP, CONICET -- Dpto. de F\'{\i}sica,
    Universidad Nacional de La Plata,\\ C.C. 67, 1900 La Plata,
    Argentina}}
\affiliation[b]{{\it Physics and Engineering Department,
    Taylor University,\\ 236 West Reade Ave., Upland, IN 46989, USA }}

\emailAdd{mileo@fisica.unlp.edu.ar}
\emailAdd{knkiers@taylor.edu}
\emailAdd{szynkman@fisica.unlp.edu.ar}
\emailAdd{dkcrane@mtu.edu}
\emailAdd{ethan\_gegner@taylor.edu}

\abstract{\hspace*{2mm}\par We present a collection of $\mathrm{CP}$-odd observables
for the process $\ppprocess$ that are linearly dependent on the scalar
($\kp$) and pseudoscalar ($\kpt$) top-Higgs coupling and hence
sensitive to the corresponding relative sign. The proposed observables
are based on triple product (TP) correlations that we extract
from the expression for the differential cross section in terms of the
spin vectors of the top and antitop quarks. In order to explore other
possibilities, we progressively modify these TPs, first
by combining them, and then by replacing the spin vectors by the
lepton momenta or the $t$ and $\tbar$ momenta by their visible parts.
We generate Monte Carlo data sets for several benchmark scenarios,
including the Standard Model ($\kp=1$, $\kpt=0$) and
two scenarios with mixed $\mathrm{CP}$ properties ($\kp=1$, $\kpt=\pm 1$).
Assuming an integrated luminosity that is consistent with that
envisioned for the High Luminosity Large Hadron Collider, using Monte Carlo-truth
and taking into account
only statistical uncertainties,
we find that the most
promising observable can disentangle the
``$\mathrm{CP}$-mixed'' scenarios 
with an effective separation of $\sim 19\sigma$. In
the case of observables that do not require the reconstruction of the $t$
and $\tbar$ momenta, the power of discrimination is up to $\sim
13\sigma$ for the same number of events. We also show that the
most promising observables can still disentangle the
$\mathrm{CP}$-mixed scenarios when the number of events
is reduced to values consistent with expectations
for the Large Hadron Collider in the near term.}

\begin{document} 
\maketitle
\flushbottom

\section{Introduction}
\label{sec1}
After the discovery of a new boson $H$ by the ATLAS \cite{atlasH} and
CMS \cite{cmsH} collaborations, it has become of crucial importance
 to determine its physical
 properties with the highest possible precision.
 The study of the new boson's couplings
 to fermions is of great relevance
 and will allow us to better understand this particle's $\mathrm{CP}$-transformation
   properties, as well as the
extent to which this particle is consistent with the Higgs boson
predicted by the Standard Model (SM) of particle physics.
It is of particular importance to test the coupling of the putative 
Higgs boson to
the top quark.  This coupling governs
the main Higgs boson production mechanism (which proceeds via gluon fusion)
and it contributes to the
important Higgs boson decay mode to two photons.
It is also
involved in the scalar-field naturalness problem -- giving rise to the
leading dependence on the cut-off energy scale in the corrections to
the Higgs mass -- and it may play an important role in the mechanism for electroweak symmetry breaking.\par

Given that the main Higgs boson production
process is dominated by a top quark loop and that
the diphoton and digluon decay channels are also mediated by a top
loop, these processes provide constraints on the scalar and
pseudoscalar $tH$ couplings, $\kp$ and $\kpt$
\cite{constraints1,constraints2,constraints3,constraints4}.
     However,
these constraints assume that there are no other sources contributing
to the corresponding effective couplings; furthermore, in the case of the
diphoton decay channel (which also involves a $W$ boson loop), it is also
assumed that the
coupling of the Higgs boson to the $W$ is standard. In this sense, the
constraints derived from measurements of Higgs boson production and decay rates
 are indirect constraints. Electric dipole moments can also
impose stringent indirect constraints on $\kpt$ by assuming that there
are no new physics (NP) particles contributing to the loops of the
relevant diagrams and in the case of the EDM of the electron that the 
electron-Higgs coupling is that
predicted by the SM \cite{constraints1,edm,cirigliano}. In order to probe
the $tH$ coupling directly, processes with smaller cross sections need
to be considered.  \par

In contrast to the $\tau H$ coupling, which can be studied through the
decay $H\rightarrow \tau^+\tau^-$ \cite{tau},
the $tH$ coupling can only be tested directly via production processes, since the
Higgs boson is kinematically forbidden from decaying to a $t\tbar$ pair.
Two types of processes are of particular interest in this regard -- the
production of a Higgs boson in association with a $t\tbar$ pair
and in association with a single top or antitop. The cross section
for associated Higgs production with a single top (antitop)
is smaller than that for production with a $t\tbar$ pair, and involves
the interference between a diagram in which the Higgs is radiated from
the top (antitop) leg and one with the Higgs emitted from the
intermediate virtual $W$ boson. Interestingly, this implies that the contraints
on $\kp$ and $\kpt$ derived from $tH$ and $\tbar H$ production are
dependent on the assumption made regarding the coupling of the Higgs boson to the
$W$ gauge boson, $\kappa_W$. Nevertheless, it is important to note that the
interference between the above mentioned diagrams
can be exploited to determine the
relative sign between $\kp$ and $\kappa_W$ (see for example
ref.~\cite{tHmaltoni,Maltoni}).  Associated Higgs production with a
$t\tbar$ pair has been studied by several authors, and
various observables sensitive to the couplings $\kp$ and $\kpt$ have
been proposed. Examples of such observables (all of which are
$\mathrm{CP}$-even) are the cross section, 
invariant mass distributions, the transverse Higgs momentum
distribution and the azimuthal angular separation between the $t$ and $\tbar$,
to name a few \cite{Guadagnoli,Boosted,Li,Golden,Khatibi,Demartin,Kobakhidze,Kobakhidze2,BhupalDev,Hagiwara}. Also, an approach based on weighted
moments and optimal observables has been developed in ref.~\cite
{Gunion1,Gunion2,Gunion3,Gunion} to discriminate the hypothesis of
a $\mathrm{CP}$-even Higgs from that of a $\mathrm{CP}$-mixed state
within the context of an $e^+ e^-$ as well as a $pp$ collider.
 Now, $\mathrm{CP}$-even observables are not
sensitive to the relative sign between the scalar and pseudoscalar
couplings $\kp$ and $\kpt$. Such observables are quadratically
dependent on these couplings and thus only provide an indirect measure
of $\mathrm{CP}$ violation. In order to be sensitive to the relative
sign between $\kp$ and $\kpt$, $\mathrm{CP}$-odd observables must be
considered. \par

Since the top quark decays before it can hadronize, its spin
information is passed on to the angular distributions of its decay
products in such a way that these particles work as spin analyzers.
As is well known, in the case of semileptonic top decay,
the charged lepton is the most powerful in this regard.
It is also known that the top quark and antiquark spins
are highly correlated in $t\tbar$ production,
a feature that
is manifested in the double angular distributions of the decay
products of the $t$ and $\tbar$ systems \cite{Mahlon1,Mahlon2,Mahlon3,atwood}.
%The spin correlations
%between the $t$ and $\tbar$ are
%dependent in turn on the $t\tbar$ production mechanism, and in
In the case of
$t\tbar H$ associated production, the $t\tbar$ spin correlations
are also sensitive to the manner in which
the top couples to the Higgs boson. 
In fact, observables that exploit
the differences in the $t\tbar$ spin configurations were used in
ref.~\cite{Biswas} to improve the discrimination of the $t\tbar H$ signal
from the dominant irreducible background $t\tbar b\bbar$, which does not
involve the Higgs boson. 

In this paper, we define a set of observables that are linearly dependent on
$\kp$ and $\kpt$ and are thus sensitive to the relative sign of these
couplings. The proposed observables are based on a particular set of
triple product (TP) correlations that we extract from the
expression for the differential cross section for $\ppprocess$,
making use of the fact that the $t$ and $\tbar$ decay products
contain spin information and are sensitive to the
nature of the $tH$ coupling, as noted above.
By using spinor techniques we relate the top
and antitop spin vectors to final state particle
momenta and separate the production process from the decay. This
allows for the straightforward identification of the contributions
that are linearly
sensitive to the couplings.
TP correlations in these
contributions incorporate the $t$ and $\tbar$ spin vectors; starting
with these TPs, we not only recover the observables given in
refs.~\cite{Ellis,Guadagnoli} but also propose additional possibilities that
have an increased sensitivity to the $tH$ coupling.
In order to establish a hierarchy in the
sensitivity of the TPs under analysis we use simulated events
to investigate three different
types of observables: asymmetries, mean
values and angular distributions. We note that TP correlations have
been used in ref.~\cite{Valencia1,Valencia2} in the context of top-quark
production and decay and in ref.~\cite{Valencia3} in the framework of
anomalous color dipole operators.  \par

The remainder of this paper is organized as follows. In
section~\ref{sec2} we study the theoretical framework for the process
$\ppprocess$ and derive a general expression for the differential
cross section.  A first set of TP correlations is then
extracted from this expression.
In section~\ref{sec3} we probe the sensitivity of these TPs to
the $tH$ coupling by using various $\mathrm{CP}$-odd
observables. Subsequent sections are dedicated to the analysis of
other $\mathrm{CP}$-odd observables. In particular,
observables based on TPs that do not contain the $t$ and $\tbar$
spin vectors, and
in certain cases incorporate the Higgs momentum, are
discussed in section~\ref{sec4}; observables that do not
involve the $t$ and $\tbar$ momenta are studied in
section~\ref{sec5}.  In section~\ref{sec6} we discuss
the experimental feasibility of the most promising observables
encountered here. The main conclusions are summarized in
section~\ref{sec7}.

%%%%%%%%%%%%%%%%%%%%%%%%%%%%%%%%%%%%%%%%%%%%%%%%%
%\newpage
%%%%%%%%%%%%%%%% Quitarlo %%%%%%%%%%%%%%%%%%%%%%%
%\setlength{\abovedisplayskip}{10.6pt}
%\setlength{\belowdisplayskip}{10.6pt}
%%%%%%%%%%%%%%%%%%%%%%%%%%%%%%%%%%%%%%%%%%%%%%%%%
%\section{Process \MakeLowercase{{\boldmath $pp\to t \bar{t}$}}$H\to$\MakeLowercase{{\boldmath $\left(b l^+ \nu_l \right) \left(\bar{b} l^-\bar{\nu}_l\right)$}}$H$. Theoretical framework}
\section{Theoretical framework for {\boldmath $pp\to t(\to b {\ell}^+ \nu_{\ell})\,\bar{t}(\to\bar{b} {\ell}^-\bar{\nu}_{\ell})\,H$}}
\label{sec2}

\begin{center}
\begin{figure}[H]
\centering
%\hspace*{-0.4cm}
\subfloat{\includegraphics[scale=0.45]{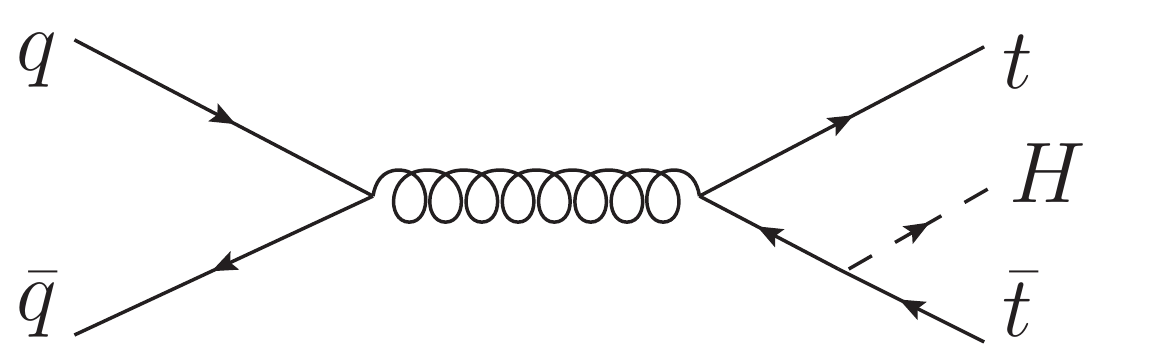}}
\hspace*{0.05\textwidth}
%\label{fig1a}}
\subfloat{\includegraphics[scale=0.45]{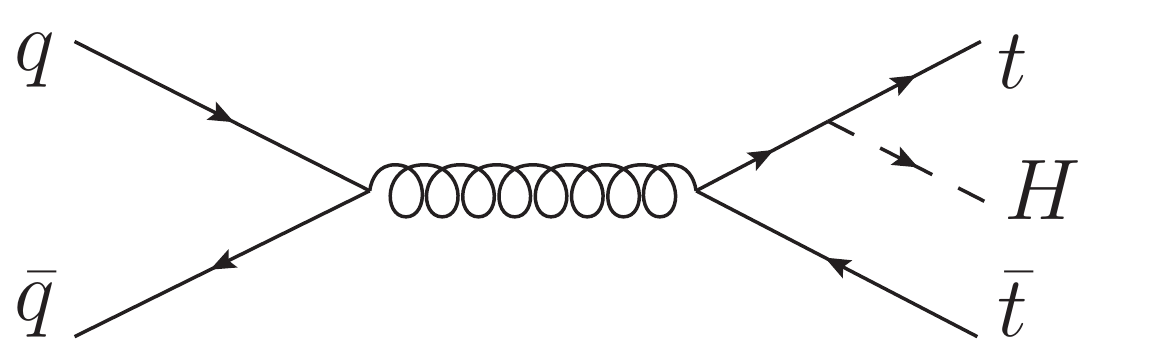}}
%\label{fig1b}}
\\[0.032\textwidth]
\subfloat{\includegraphics[scale=0.45]{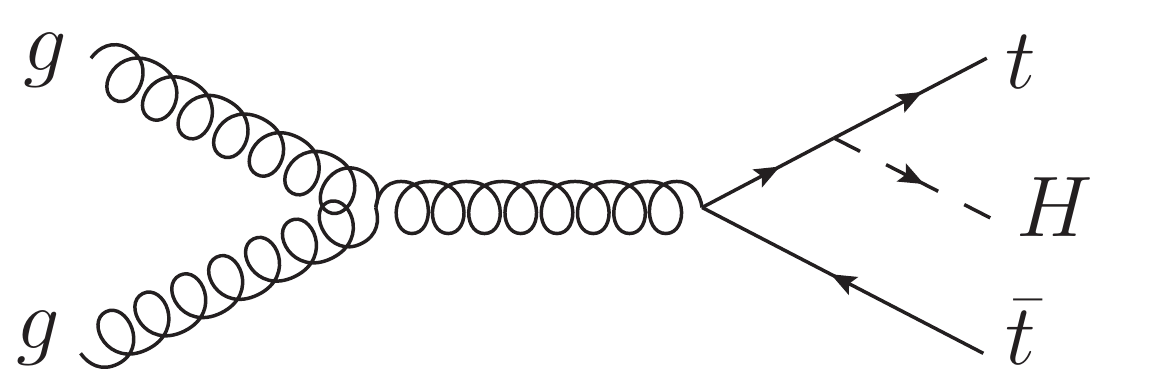}}
\hspace*{0.05\textwidth}
%\label{fig1b}}
\subfloat{\includegraphics[scale=0.45]{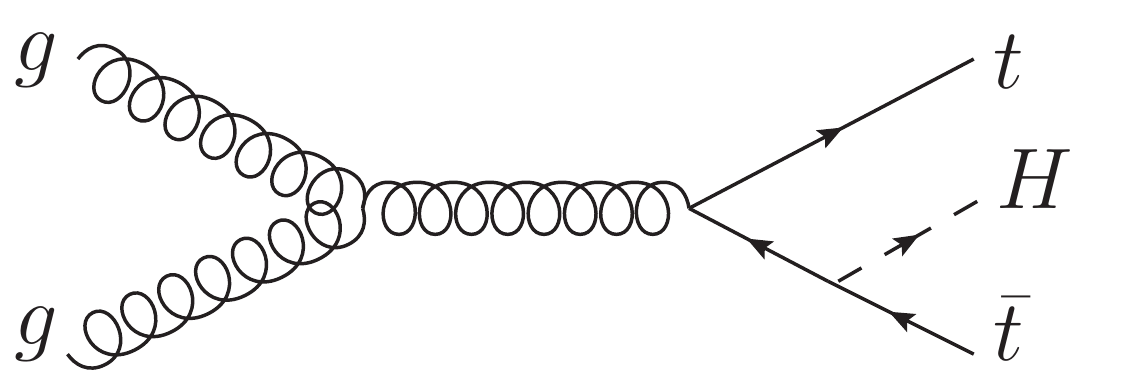}}
%\label{fig1c}}
\\[0.032\textwidth]
\subfloat{\includegraphics[scale=0.45]{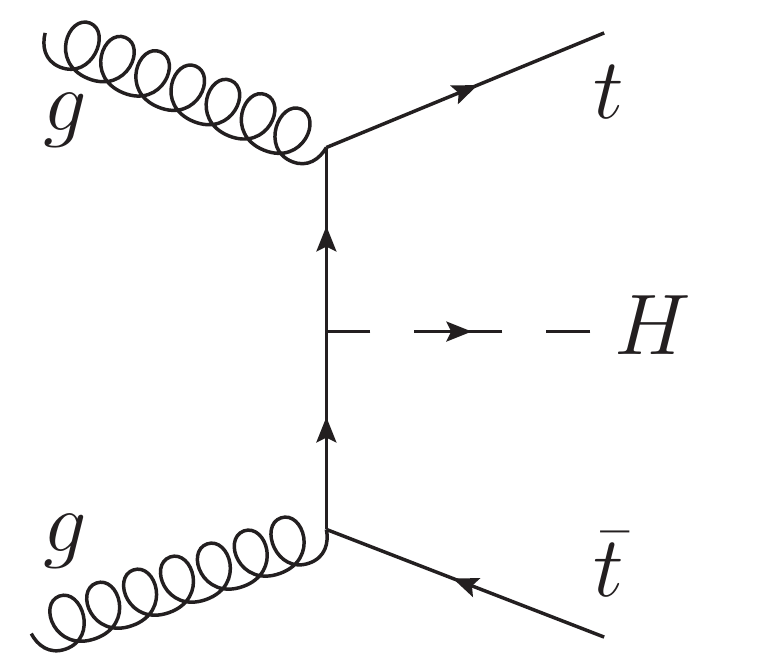}}
%\label{fig1b}}
\hspace*{0.025\textwidth}
\subfloat{\includegraphics[scale=0.45]{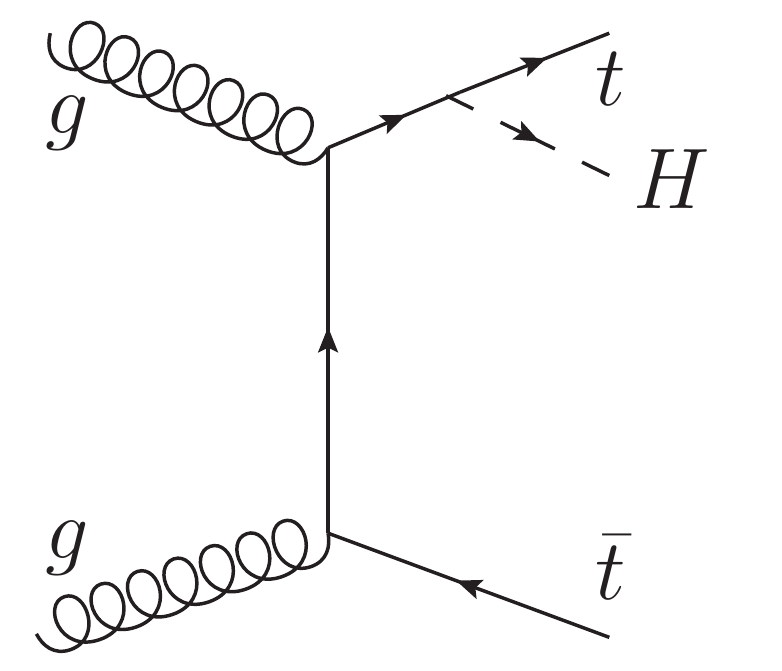}}
%\label{fig1c}}
\hspace*{0.025\textwidth}
\subfloat{\includegraphics[scale=0.45]{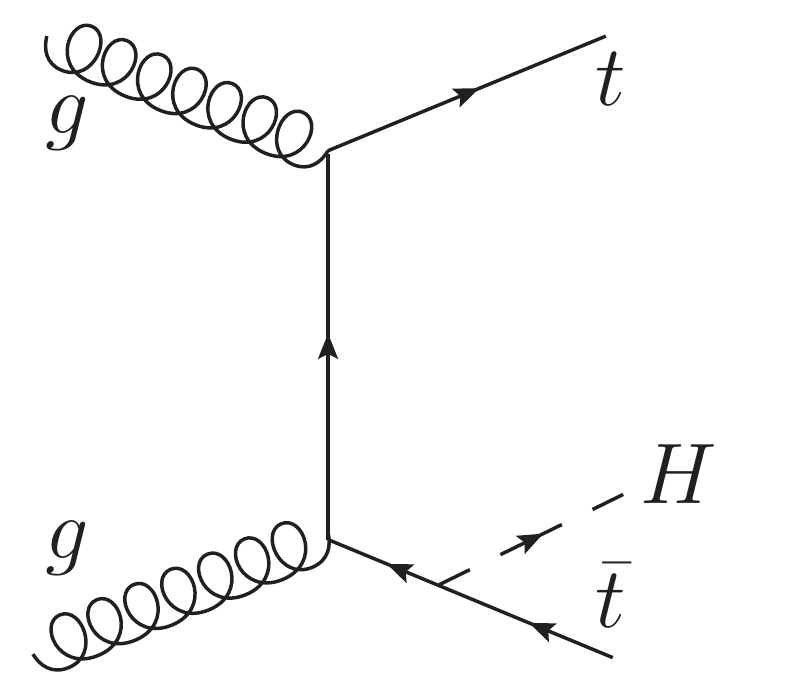}}
%\label{fig1b}}
\vspace*{0.02\textwidth}
\caption{Tree-level Feynman diagrams contributing to $t\tbar H$ production
  at the LHC. Three more diagrams are obtained by exchanging the gluon
  lines in the $t$-channel diagrams.}
\label{fig1}
\end{figure}
\end{center}

At the Large Hadron Collider (LHC) $t\bar{t}H$ production proceeds via $q\bar{q}$
annihilation and $gg$ fusion processes. The relevant leading-order Feynman
diagrams are displayed in figure~\ref{fig1}, where the first two
rows show the $q\bar{q}$ and $gg$ $s$-channel diagrams, and
the last one depicts the $gg$ $t$-channel diagrams.  Three more
$gg$-initiated diagrams are obtained by exchanging the gluon
lines in the third row.
We describe the $tH$ coupling with the effective Lagrangian
\beq
\label{eq1}
\mathcal{L}_{t\bar{t}H}=-\frac{m_t}{v}(\kp \tbar t+i\kpt \tbar
\gamma_5 t)H,
\eeq
where $v=246~\mathrm{GeV}$ is the SM Higgs vacuum
expectation value, and the coefficients $\kp$ and $\kpt$ parameterize 
the scalar and pseudoscalar interaction, respectively. The
SM case is obtained for $\kp=1$ and $\kpt=0$, while the values $\kp=0$
and $\kpt\neq 0$ parameterize a $\mathrm{CP}$-odd Higgs boson.

Before turning to a discussion of $\mathrm{CP}$-odd observables,
it is useful to consider a few theoretical aspects of
the process $\ppprocess$, in which the top and antitop both decay
semileptonically.
In the following subsections we derive a ``factorized''
expression for the gluon fusion contribution to this process
and then use this expression
to isolate various mathematical quantities
that will be useful as we construct $\mathrm{CP}$-odd observables.\par

%%%%%%%%%%%%%%%%%%%% Figura 1%%%%%%%%%%%%%%%%%%%%%%%%%%%%%
%/home/nico/jaxodraw-2.1-0/
\vspace*{-1cm}
%%%%%%%%%%%%%%%%%%%%%%%%%%%%%%%%%%%%%%%%%%%%%%%%%%%%%%%%%%%%
\subsection{Factorized expression for the scattering cross section}
\label{subsec:factorize}

In this subsection we focus on the $gg$-initiated
contributions
to $t\tbar H$ production, since these dominate over the
the quark-antiquark annihilation contributions. As we shall show below, assuming the
narrow width approximation for the top and antitop quarks, the unpolarized
differential cross section for $gg\to t(\to
b{\ell}^+\nu_{\ell})\,\tbar(\to \bbar {\ell}^- \nubar)\,H$ may be
written in the following ``factorized'' form,\footnote{The reader
  is referred to the discussion following eq.~(\ref{eq17}) for some qualifying
  remarks regarding the ``factorization'' of this expression.}
\beq
\label{eq2}
%dd\sigma(gg\to (bl^+\nu_l)(\bbar l^- \nubar) H)=\sum_{\substack{bl^+\nu_l \\ \tiny{\mathrm{spins}}}}\,\sum_{\substack{\bbar l^-\nubar \\ \mathrm{\tiny{spins}}}}\left(\frac{2}{\Gamma_t}\right)^2\,d\sigma(gg\to t(n_t)\tbar (n_{\tbar})H)\,d\Gamma(t(n_t)\to bl^+\nu_l)\,d\Gamma(\tbar (n_{\tbar})\to \bbar l^-\nubar)
d\sigma =\sum_{\substack{b{\ell}^+\nu_l \\ \tiny{\mathrm{spins}}}}\,
   \sum_{\substack{\bbar {\ell}^-\nubar \\ \mathrm{\tiny{spins}}}}\left(\frac{2}{\Gamma_t}\right)^2\,
   d\sigma(gg\to t(n_t)\tbar (n_{\tbar})H)\,
   d\Gamma(t\to b{\ell}^+\nu_{\ell})\,
   d\Gamma(\tbar \to \bbar {\ell}^-\nubar),
\eeq  
where $d\sigma(gg\to t(n_t)\tbar (n_{\tbar})H)$ is the differential
cross section for the production of a top and antitop quark,
with spin vectors $n_t$ and $n_{\bar{t}}$, respectively, along with a Higgs
boson.  Also, $d\Gamma(t\to b{\ell}^+\nu_{\ell})$ and
$d\Gamma(\tbar \to \bbar {\ell}^-\nubar)$ are the
partial differential decay widths for an unpolarized top and
anti-top quark.  The four-vectors
$n_t$ and $n_{\tbar}$ are not arbitrary, but are
given by particular combinations of the
momenta of the $t,\tbar, \ell^+$ and $\ell^-$~\cite{Arens},
\bea
\label{eq3}
n_t&=&-\frac{p_t}{m_t}+\frac{m_t}{(p_t\cdot p_{{\ell}^+})}p_{{\ell}^+}\\
\label{eq4}
n_{\tbar}&=&\,\frac{p_{\tbar}}{m_t}-\frac{m_t}{(p_{\tbar}\cdot p_{{\ell}^-})}p_{{\ell}^-}.
\eea
Expressions similar to eq.~(\ref{eq2}) have been derived previously for the
production of short-lived particles in $e^-e^+$ colliders
\cite{kawasaki} and for $t\tbar$ production both in
$e^-e^+$ colliders \cite{Arens} and $pp$ colliders
\cite{ale1,ale2,ale3,ale4}. \par
%%%%%%%%%%%%%%%%% FIGURA 2 %%%%%%%%%%%%%%%%%%%%%%
\begin{center}
\begin{figure}[H]
\centering
\includegraphics[scale=0.6]{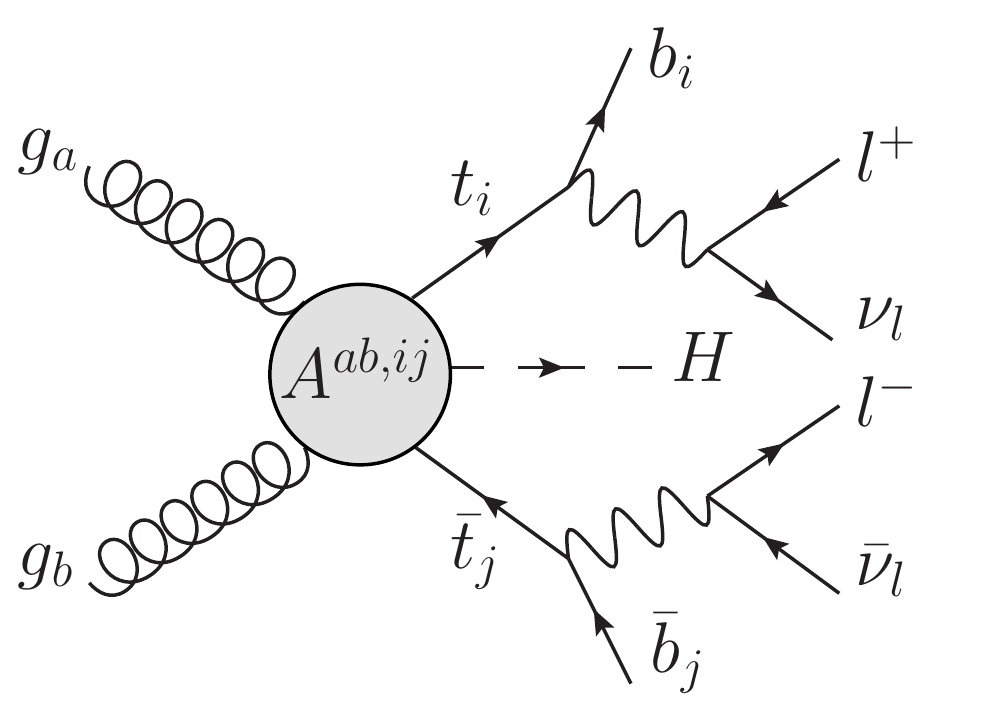}
\vspace*{0.02\textwidth}
\caption{Schematic representation of the process $g_ag_b\to t(\to
  b_i{\ell}^+\nu_{\ell})\tbar(\to \bbar_j {\ell}^- \nubar) H$. The
  indices $i,j$ denote the colours of the quarks, while $a,b$ are gluon
  indices.}
\label{fig2}
\end{figure}
\end{center}
%%%%%%%%%%%%%%%%%%%%%%%%%%%%%%%%%%%%%%%%%%%%%%%%%
\vspace*{-0.85cm}
To derive the above expressions, we begin by considering the
schematic representation for the process $g_ag_b\to t(\to
b_i{\ell}^+\nu_{\ell})\,\tbar(\to \bbar_j {\ell}^- \nubar)\,H$
that is sketched in figure~\ref{fig2}.
Here $a$ and $b$ denote the initial-state gluons and
$i$ and $j$ refer to the colours of the top and antitop quarks.
%\par
%
The amplitude for this process may be written in the following
compact form
\beq
\label{eq6}
\mathcal{M}^{ab,ij}=\bar{\psi}_t\,\mathcal{A}^{ab,ij}\,\psi_{\tbar}\,,
\eeq
where the spinors $\bar{\psi}_t$ and $\psi_{\tbar}$
contain all of the information regarding the decay of the
virtual top and anti-top, respectively,
and where the quantity $A^{ab,ij}$ is given by
\beq
\label{eq5}
\mathcal{A}^{ab,ij}\equiv A^{ab,ij}_{\mu\nu}(\epsilon_{\lambda_a})^{\mu}(\epsilon_{\lambda_b})^{\nu}=\sum_{k=1}^8 \mathcal{A}^{ab,ij}_k =\kp \sum_{k=1}^8 \mathcal{S}^{ab,ij}_k + i\kpt \sum_{k=1}^8 \mathcal{P}^{ab,ij}_k.
\eeq 
The sum over $k$ in the above expression
corresponds to the eight gluon-initiated
diagrams indicated in figure~\ref{fig1}; also,
$\epsilon_{\lambda_a}$ and $\epsilon_{\lambda_b}$ are the
polarization vectors corresponding to $g_a$ and $g_b$, respectively.
In the last equality in eq.~(\ref{eq5})
we have explicitly separated the amplitude into
two sums, with one sum corresponding to the scalar contributions
and the other to the pseudoscalar ones.
Taking all of the final-state particles to be massless, we can use the
spinor techniques
developed in ref.~\cite{Kleiss} to write $\bar{\psi}_t$ and
$\psi_{\tbar}$ as follows\footnote{These spinor techniques can also be used
  for massive final-state particles. Given the
  energy scale involved in the process in question, however, the assumption
  of massless final-state particles is sensible and greatly
  simplifies the derivation of eq.~(\ref{eq2}).}
\beq
\label{eq7}
\bar{\psi}_t = -g^2\, \mathbb{P}_t(t)\,\mathbb{P}_W(t-b)\,
   \langle b-|\nu_{\ell}+\rangle \langle {\ell}^+\!+|(\tslash+m_t)
\eeq
\beq
\label{eq8}
\psi_{\tbar} = g^2\, \mathbb{P}_t(\tbar)\,\mathbb{P}_W(\tbar-\bbar)\,
   \langle\nubar+|\bbar-\rangle (\tbslash-m_t)|{\ell}^-+\rangle,\\[1mm]
\eeq
%\bea
%\label{eq7}
%\bar{\psi}_t &=& -g^2\, \mathbb{P}_t(t)\,\mathbb{P}_W(t-b)\,
%   \langle b-|\nu_{\ell}+\rangle \langle {\ell}^+\!+|(\tslash+m_t)\\[0.4cm]
%\label{eq8}
%\psi_{\tbar} &=& g^2\, \mathbb{P}_t(\tbar)\,\mathbb{P}_W(\tbar-\bbar)\,
%   \langle\nubar+|\bbar-\rangle (\tbslash-m_t)|{\ell}^-+\rangle,
%   \vspace*{0.8cm}
%\eea
%
where $|i+(-)\rangle \equiv (1/2)(1\pm \gamma^5)\,\psi_i$
represents a right-handed (left-handed) chiral spinor for final-state
particle $i$ and $\langle i+(-)|$ represents the corresponding adjoint spinor.
Also, $\mathbb{P}_t(q)=(q^2-m^2_t+im_t\Gamma_t)^{-1}$ and
$\mathbb{P}_W(q)=(q^2-m^2_W+im_W\Gamma_W)^{-1}$, and
we have denoted the momenta of the various particles by the symbols
that refer to the names of those particles~\cite{Mangano}.

Using the expressions defined above for $\bar{\psi}_t$ and
$\psi_{\tbar}$, we can write the amplitude $\mathcal{M}^{ab,ij}$
in a form that is (in a sense) factorized.
As a first step, we insert
eqs.~(\ref{eq7}) and (\ref{eq8}) into eq.~(\ref{eq6}), yielding
\beq
\label{eq9}
\mathcal{M}^{ab,ij}=\!-g^4\,\mathbb{P}_t(t)\,\mathbb{P}_t(\tbar)\,\mathbb{P}_W(t-b)\,\mathbb{P}_W(\tbar-\bbar)\,\langle b-|\nu_{\ell}+\rangle \langle\nubar+|\bbar-\rangle \sqrt{\strut2(t\cdot {\ell}^+)}\sqrt{\strut2(\tbar\cdot {\ell}^-)}\,[\bar{\phi}_t \mathcal{A}^{ab,ij}\phi_{\tbar}],
%\frac{1}{(t^2-m^2_t+im_t\Gamma_t)}\,\frac{1}{(\tbar^2-m^2_t+im_t\Gamma_t)}\,\frac{1}{((t-b)^2-M^2_W)}\frac{1}{((\tbar-\bbar)^2-M^2_W)}\,\langle b-|\nu+\rangle\,\langle\nubar+|\bbar-\rangle
\eeq
where the spinors $\phi_{t}$ and $\phi_{\tbar}$ are defined as
\beq
\label{eq10}
\phi_t=\frac{(\tslash +m_t)}{\sqrt{\strut2(t\cdot {\ell}^+)}}|{\ell}^++\rangle
\eeq
\beq
\label{eq11}
\phi_{\tbar}=\frac{(\tbslash -m_t)}{\sqrt{\strut2(\tbar\cdot {\ell}^-)}}|{\ell}^-+\rangle \,.
\eeq
Note that in writing down the above expressions we have adopted the narrow-width
approximation for the top and antitop quarks.\footnote{Since eq.~(\ref{eq9}) contains the top quark propagator
  term
  $\mathbb{P}_t(t)$, for example, $|\mathcal{M}^{ab,ij}|^2$ contains
  the factor $((t^2-m^2_t)^2+m^2_t\Gamma^2_t)^{-1}$,
  which is replaced by $(\pi/(m_t\Gamma_t))\delta(t^2-m^2_t)$
  in the narrow-width approximation.  Thus, except for the propagator terms $\mathbb{P}_t(t)$ and $\mathbb{P}_t(\tbar)$,
   we take the four-vector $t$ appearing in eqs.~(\ref{eq9})-(\ref{eq11}) to be
  on shell, satisfying $t^2 = m_t^2$.} 
Working out the projection operators $\phi_t\,\bar{\phi}_t$
and $\phi_{\tbar}\,\bar{\phi}_{\tbar}$, we have
\beq
\label{eq12}
\phi_t\,\bar{\phi}_t=\frac{1}{2}(1+\nslash_{\!t}\gamma^5)(\tslash +m_t)
\eeq
and
\beq
\label{eq13}
\phi_{\tbar}\,\bar{\phi}_{\tbar}=\frac{1}{2}(1+\nslash_{\!\tbar}\gamma^5)(\tbslash -m_t),
\eeq
with $n_t$ and $n_{\tbar}$ being the four-vectors defined in
eqs.~(\ref{eq3}) and (\ref{eq4}).  Thus, $\phi_t$ and $\phi_{\tbar}$
may be regarded as
describing a top quark with spin vector $n_t$ and an antitop quark
with spin vector $n_{\tbar}$, respectively.

As a final step toward factorizing the amplitude $\mathcal{M}^{ab,ij}$,
we note that the amplitude for a top quark with spin vector
$n_t$ to decay into $b{\ell}^+\nu_{\ell}$ is given by
\beq
\label{eq14}
\mathcal{M}(t(n_t)\to b{\ell}^+\nu_{\ell})=ig^2\mathbb{P}_W(t-b)\langle b-|\nu_{\ell}+\rangle\sqrt{\strut2(t\cdot {\ell}^+)} \,,
\eeq
and likewise,
\beq
\label{eq15}
\mathcal{M}(\tbar(n_{\tbar})\to \bar{b}{\ell}^-\bar{\nu}_{\ell})=ig^2\mathbb{P}_W(\tbar-\bar{b})\langle \bar{\nu}_{\ell}+|\bar{b}-\rangle\sqrt{\strut2(\tbar\cdot {\ell}^-)}.
\eeq
Furthermore, the term inside the square brackets in
eq.~(\ref{eq9}) is the amplitude for producing a top quark
with spin vector $n_t$, along with an anti-top with spin vector
$n_{\tbar}$ and a Higgs boson,
\beq
\label{eq16}
\mathcal{M}(g_ag_b \to t^i(n_t)\tbar^j(n_{\tbar})H)=\bar{\phi}_t \mathcal{A}^{ab,ij}\phi_{\tbar}.
\eeq
Combining eqs.~(\ref{eq14})-(\ref{eq16}), we can write eq.~(\ref{eq9})
in a form that appears to be factorized,
\beq
\label{eq17}
\mathcal{M}^{ab,ij}=\mathbb{P}_t(t)\mathbb{P}_t(\tbar)\,\mathcal{M}(t(n_t)\to b{\ell}^+\nu_{\ell})\,\mathcal{M}(\tbar(n_{\tbar})\to \bar{b}{\ell}^-\bar{\nu}_{\ell})\,\mathcal{M}(g_ag_b \to t^i(n_t)\tbar^j(n_{\tbar})H) \,.
\eeq
It is important
to note that, even though the above expression has the appearance of being
factorized into production and decay parts, this apparent
factorization is a bit misleading.  In particular, the amplitude for
$t\tbar H$ production contains the top and antitop quark
spin four-vectors $n_t$ and $n_{\tbar}$, which
depend on final-state kinematical quantities (see
eqs.~(\ref{eq3}) and (\ref{eq4})).
With this qualification in mind, we may now use the
amplitude in eq.~(\ref{eq17})
to determine the corresponding scattering cross section.
After some manipulation of the phase space variables
to take advantage of the presence of the
propagator terms, $\mathbb{P}_t(t)$ and $\mathbb{P}_t(\tbar)$,
we arrive at the expression in eq.~(\ref{eq2}).\footnote{The reader may note that in the differential widths of $t\to b{\ell}^+\nu_{\ell}$ and $\tbar\to \bar{b}{\ell}^-\bar{\nu}_{\ell}$ appearing in 
eq.~(\ref{eq2}), the spin states of the top and antitop have been averaged. Interestingly, under the assumption of massless final-state particles, the amplitudes $\mathcal{M}(t(-n_t)\to b{\ell}^+\nu_{\ell})$ and $\mathcal{M}(\tbar(-n_{\tbar})\to \bar{b}{\ell}^-\bar{\nu}_{\ell})$ vanish. } This expression
also has the appearance of being factorized, but
qualifying remarks, similar to those above, apply.

%%%%%%%%%%%%%%%%%%% Quitarlo %%%%%%%%%%%%%%%%%%%%
\setlength{\abovedisplayskip}{10.2pt}
\setlength{\belowdisplayskip}{10.2pt}
%%%%%%%%%%%%%%%%%%%%%%%%%%%%%%%%%%%%%%%%%%%%%%%%%

\subsection{Origin of triple product terms}
\label{subsec:origin}

The expression derived above for the scattering cross
section (see eq.~(\ref{eq2}), as well as eq.~(\ref{eq17}))
provides significant insight into
how one might analyze $\ppprocess$ in order to determine
the nature of the top-Higgs coupling.  In particular,
let us focus on the production amplitude,
$\mathcal{M}(g_ag_b \to t^i(n_t)\tbar^j(n_{\tbar})H))$, which forms
part of the overall amplitude in eq.~(\ref{eq17}).
The absolute value squared of the production amplitude
is used to determine $d\sigma(gg\to t(n_t)\tbar (n_{\tbar})H)$,
which in turn forms part of the expression for the ``factorized'' cross section
in eq.~(\ref{eq2}).  Summing over colour and gluon indices
we have
\beq
\label{eq18}
\sum_{\substack{a,b \\ i,j}}|\mathcal{M}(g_ag_b \to t^i(n_t)\tbar^j(n_{\tbar})H)|^2=\sum_{\substack{a,b \\ i,j}}\left|\sum^{8}_{k=1}C^{ab,ij}_k\,\bar{\phi}_t (\kp\mathcal{S}_k+i\kpt\mathcal{P}_k)\phi_{\tbar}\right|^2,
\eeq
where we have separated the colour structure of each diagram by defining
$\mathcal{S}^{\,ab,ij}_k= C^{ab,ij}_k \mathcal{S}_k$ and
$\mathcal{P}^{\,ab,ij}_k= C^{ab,ij}_k \mathcal{P}_k$
(see eqs.~(\ref{eq5}) and (\ref{eq16})). Also, the factors
$g^2_s m_t/v$ and $-ig^2_s m_t/v$ arising from the vertices of the $t$-
and $s$-channel diagrams, respectively, have been included in the
definition of $C^{ab,ij}_k$ for convenience. The terms linear in $\kp$
and $\kpt$ can be written as
\beq
\label{eq19}
\mathcal{O}(\kp\kpt)\to \frac{1}{2}\kp\kpt \sum_{k,r}\mathbb{C}_{kr}\mathrm{Im}
\left\lbrace \mathrm{Tr}\left[ (1+\nslash_t \gamma^5)(\tslash+m_t)\mathcal{S}_k(1+\nslash_{\tbar}\gamma^5
 )(\tbslash -m_t)\tilde{\mathcal{P}}_r \right] \right\rbrace ,
\eeq
where the factor $\mathbb{C}_{kr}=\sum_{ab,ij}C^{ab,ij}_k
C^{ab,ij*}_r$ is real and where $\tilde{\mathcal{P}}_r = \gamma^0
\mathcal{P}^{\dagger}_r \gamma^0$.
The only terms that yield non-zero contributions
in the above sum are those with an
odd number of $\gamma^5$ matrices; these lead to triple-product
(TP) correlations
of the form $\epsilon_{\alpha\beta\gamma\delta}\,p^{\alpha}_ap^{\beta}_bp^{\gamma}_cp^{\delta}_d$,
where $p_a$-$p_d$ represent various four momenta associated with the process.
In contrast,
it can be seen from eq.~(\ref{eq18}) that the terms proportional to
$\kp^2$ and $\tilde{\kappa}^2_t$ descend from traces containing
an even number of $\gamma^5$
matrices and can be written in terms of scalar products of the
available momenta.\par
%\vspace*{-2mm}

With the above considerations in mind, it is useful to
write a general expression for the differential cross
section $d\sigma(gg\to t(n_t)\tbar (n_{\tbar})H)$
in terms of the momenta $q=(q_1-q_2)/2$,
$Q=(q_1+q_2)/2$, $t$, $\bar{t}$, $n_t$ and $n_{\tbar}$, where
$q_{1,2}$ denote the momenta of the initial-state gluons. Note that with this
choice, $q\cdot Q=0$ and $Q^2=-q^2=M^2_{t\tbar H}/4$, where $M_{t\tbar
  H}$ is the invariant mass of the $t\tbar H$ system.
Fifteen TPs can be constructed from these six
four-vectors,\footnote{We note that these fifteen
  TPs are not linearly independent (see the epsilon relations
  discussed in ref.~\cite{identities}).} so that
\beq
\label{eq20}
d\sigma(gg\to t(n_t)\tbar (n_{\tbar})H)= \kp^2\,f_1(p_i\cdot p_j)+\tilde{\kappa}^2_t\,f_2(p_i\cdot p_j)+\kp\kpt\,\sum^{15}_{l=1}g_l(p_i\cdot p_j)\,\epsilon_l,
\eeq   
where
$\epsilon_l=\epsilon_{\alpha\beta\gamma\delta}\,p^{\alpha}_ap^{\beta}_bp^{\gamma}_cp^{\delta}_d$
denotes the $l$th TP (we adopt the convention $\epsilon_{0123}=+1$) and where $p_i$ and $p_j$ refer to any of the
six momenta.  The
functions $f_{1,2}$ and $g_k$ depend only on the possible scalar
products and are therefore even under a parity transformation
($\mathrm{P}$). However, the terms linear in $\kp\kpt$ are
$\mathrm{P}$-odd due to the presence of the $\mathrm{P}$-odd
TPs. Hence, only the functions $f_{1,2}$ will contribute to the total
cross section, whereas the TP terms will be sensitive to the sign of
the anomalous coupling $\kpt$. Of the fifteen TPs mentioned above,
we will focus on those that contain both of the spin
vectors $n_t$ and $n_{\tbar}$, but do not include $q$.
The decision not to consider $q$-dependent TPs is motivated by the fact
that $q$ cannot be expressed in terms of the momenta of final state
particles (as $Q$ can, by virtue of energy-momentum conservation). The
decision to focus on TPs that contain both $n_t$ and $n_{\tbar}$
is rooted in the fact that the spins of pair-produced top and antitop quarks
are highly correlated at hadron colliders 
(even though the quarks themselves are unpolarized).
Observables that combine the decay products of the
$t$ and $\tbar$ will be sensitive to this spin
correlation~\cite{Bernreuther}.  A similar behaviour is expected in $t\tbar H$
production, where it can be shown that single-spin asymmetries
vanish~\cite{Ellis,Biswas}. Hence, in order to construct observables
sensitive to the structure of the $tH$ coupling, we will restrict our attention
to those
TPs that include information on the decay products of both the top and
anti-top quarks. Only five of the fifteen TPs in eq.~(\ref{eq20}) do
not involve the four vector $q$ and, among these, only three
include both $n_t$ and $n_{\tbar}\,$.  Thus, we will restrict our attention
to the following TPs
\beq
\label{eqa1}
\epsilon_1\equiv\epsilon(t,\tbar,n_t,n_{\tbar}),\,\,
\vspace*{1mm}
\eeq
\beq
\label{eqa2}
\epsilon_2\equiv\epsilon(Q,\tbar,n_t,n_{\tbar}),
\vspace*{1mm}
\eeq
\beq
\label{eqa3}
\epsilon_3\equiv\epsilon(Q,t,n_t,n_{\tbar}).
\vspace*{1mm}
\eeq
\par  

Before turning to a consideration of various CP-odd observables,
we remark that even though all of the above discussion took place
within the context of $gg$-initiated production, similar conclusions
are obtained for $q\qbar$-initiated production. In particular, the
definitions of the spin vectors in eqs.~(\ref{eq3})-(\ref{eq4}) and
the general form of $d\sigma$ introduced in eq.~(\ref{eq20}) are valid
in both cases.
%%%%%%%%%%%%%%%%%%%%%%%%%%%%%%%%%%%%%%%%%%%%%%%%%
%\newpage
%%%%%%%%%%%%%%%%%%%%%%%%%%%%%%%%%%%%%%%%%%%%%%%%%
%\setlength{\belowdisplayskip}{10pt} \setlength{\belowdisplayshortskip}{10pt}
%\setlength{\abovedisplayskip}{10pt} \setlength{\abovedisplayshortskip}{10pt}
\bigskip
\section{$\mathrm{\mathbf{CP}}$-odd observables}
\label{sec3}
In this section we present three types of observables based on the TPs discussed
in section~\ref{sec2}, namely, asymmetries, angular
distributions and mean values. These observables are sensitive not only to the
magnitude of the pseudoscalar coupling $\kpt$, but also to its
sign.  In order to test the various observables, we have
used $\mathtt{MadGraph5\_aMC@NLO}$ \cite{Madgraph} to simulate the process
$\ppprocess$ at parton level for different values of the couplings
$\kp$ and $\kpt$.  In all cases we have generated $10^5$ events
and have assumed a center-of-mass energy of
$14\,\mathrm{TeV}$.\footnote{Note that, since we generate
  the same number of events in
  each case, the corresponding integrated luminosities are different, since the
  cross section depends on the value of $\kpt$.}
We have also imposed the
following set of cuts: $p_T$ of leptons $> 10\,\mathrm{GeV}$, $|\eta|$
of leptons $< 2.5$, $|\eta|$ of $b$ jets $< 2.5$ and $\Delta
R_{\ell\ell}>0.4$.  Note that we have used this
somewhat large number of events ($10^5$) in order to determine clearly
the extent to which the proposed observables are sensitive to the anomalous
coupling.  Section~\ref{sec6} contains an
analysis of the experimental feasibility of the more promising observables.

Before continuing on to our analysis, let us make a few comments
regarding the values that we choose for $\kp$ and $\kpt$.
First of all, we note that if the pseudoscalar
coupling $\kpt$ is the only source of physics beyond the SM,
then indirect contraints (based on the signal strength of $gg\to H \to \gamma\gamma$) 
disfavour $\kp < 0$ but do not
resolve the degeneracy in the sign of $\kpt$ \cite{Guadagnoli}. On
the other hand, if one assumes that the tensor structure of the
Higgs interactions are the
same as those of the SM and if one parameterizes these interactions via
one universal Higgs coupling to vector bosons, $\kappa_V$, and one
universal Higgs coupling to fermions, $\kappa_f$, then the measured signal
strengths provided by the ATLAS and CMS collaborations are compatible with
the values predicted by the SM, (namely, $\kappa_f=1$ and $\kappa_V=1$).
With these facts in mind, we will, for the most part,
set the value of the scalar coupling to its SM
value ($\kp=1$) and will allow the pseudoscalar coupling to take on
various values (including both possible signs). In particular, we
analyze the cases $\kpt=0,\pm 0.25, \pm 0.5, \pm 0.75,\pm 1$.
We shall often focus on the scenarios with $\kp=1$ and
$\kpt=\pm 1$, which we shall refer to as the ``$\mathrm{CP}$-mixed''
scenarios.
In addition, we also provide some analysis regarding the pure
$\mathrm{CP}$-odd case ($\kp=0,\kpt=1$).
%\newpage
%%%%%%%%%%%%%%%%%%%%%%%%%%%%%%%%%%%%%%%%%%%%%%%%%
\subsection{Asymmetry}
%\bigskip
\label{sec3.1}

The first type of $\mathrm{CP}$-odd observable that we will consider is an
asymmetry that compares the number of events
for which a given TP is positive to that for which it is negative.
Normalizing to the total number of events, we define
\beq
\label{eq21}
\mathcal{A}(\epsilon)=\frac{N(\epsilon > 0)-N(\epsilon < 0)}{N(\epsilon > 0)+N(\epsilon < 0)}.
\eeq 
By construction, $\mathcal{A}\in [-1,+1]$.
Based on the general expression given in eq.~(\ref{eq20}), we expect
the following functional form for the asymmetry,
\beq
\label{eq22}
\mathcal{A}(\epsilon)=\frac{A\kp\kpt}{B\kappa^2_t+C\tilde{\kappa}^2_t},
\eeq 
which for $\kp=1$ can be parameterized as
\beq
\label{eq23}
\mathcal{A}(\epsilon)=\frac{a\kpt}{1+b\tilde{\kappa}^2_t},
\eeq 
where the parameter $a\equiv A/B$ determines the sensitivity to the
pseudoscalar coupling, whereas $b\equiv C/B$ quantifies the deviation
from linear behaviour.

table~\ref{table1} shows numerical results for the
asymmetries associated with three different TPs,
$\epsilon_1$, $\epsilon_2$ and $\epsilon_3$, taking
$\kp=1$ and $\kpt=0,\pm 1$.  The asymmetry $\mathcal{A}$
is shown in each case, along with $\mathcal{A}/\sigma_{\mathcal{A}}$, where
$\sigma_{\mathcal{A}}$ is the corresponding
statistical uncertainty.  As is
evident from the table, the asymmetries in question provide
a clear separation between the SM and the 
$\mathrm{CP}$-mixed cases, with typical deviations being of
order $10\sigma$.  Furthermore, the asymmetries
for the SM case are each statistically consistent with zero,
as one would expect.
The three asymmetries also allow one to
determine the sign of $\kpt$,
with the $\kpt = \pm 1$ cases effectively separated by more than $20\sigma$.
The sensitivity of
the asymmetry is quite similar for the three TPs, as can be seen by
including other values of $\kpt$ and using the expression in
eq.~(\ref{eq23}) as a fitting function (see figure~\ref{fig3}).
Performing such a fit, 
we obtain $(a=-0.057\pm 0.006, b=0.5\pm 0.2),
(a=-0.056\pm 0.006, b=0.5 \pm 0.2)$ and $(a=0.058\pm 0.006, b=0.6 \pm
0.2)$ for $\epsilon_1$, $\epsilon_2$ and $\epsilon_3$,
respectively.
\newcolumntype{C}[1]{>{\centering\arraybackslash}p{#1}}
\renewcommand{\arraystretch}{1.4}
\begin{table}[H]
  \caption{Asymmetries for three different scenarios,
    obtained by using $10^5$ simulated events, for the
    TPs $\epsilon_1=\TPa, \epsilon_2=\TPb$ and $\epsilon_3=\TPc$.
    The three scenarios
  correspond to the SM ($\kp=1$ and $\kpt= 0$) and the
  two ``$\mathrm{CP}$-mixed'' cases (defined by
  $\kp=1$ and $\kpt=\pm 1$).}
\label{table1}
\begin{center}
\begin{tabular}{|C{0.8cm}|C{0.8cm}||C{1.9cm}|C{1.9cm}||C{1.9cm}|C{1.9cm}||C{1.9cm}|C{1.9cm}|}
%\begin{tabular}{|c|r||r|c||r|c||r|c|}
\hhline{|========|}
%\hhline{|--------|}
$\kappa_t$&$\tilde{\kappa}_t$~~&$\mathcal{A}(\epsilon_1)$~~&$\mathcal{A}(\epsilon_1)/\sigma_{\mathcal{A}}$& $\mathcal{A}(\epsilon_2)$~~&$\mathcal{A}(\epsilon_2)/\sigma_{\mathcal{A}}$&$\mathcal{A}(\epsilon_3)$~~&$\mathcal{A}(\epsilon_3)/\sigma_{\mathcal{A}}$  \\ 
\hhline{|========|} 
%\hhline{|--------|}
$1$ & $-1$~~~ & $0.0315$~ & $10.0$ & $0.0332$~ & $10.5$~ & $-0.0307$~~~ & $-9.7$~~~\\[0.6mm]
\hline
$1$ & $0$ & $-0.0021$~~~ & $-0.7$~~~ & $0.0009$~ & $0.3$~ & $-0.0011$~~~ & $-0.3$~~~\\[0.6mm]
\hline
$1$ & $1$ & $-0.0379$~~~ & $-12.0$~~~ & $-0.0411$~~~& $-13.0$~~~ & $\,0.0378$~ & $12.0$ \\[0.6mm]
\hhline{|========|}
%\hhline{|--------|}
\end{tabular}
\end{center} 
\end{table}
%%%%%%%%%%%%%%%%%%%%%%%%% FIGURA 3 %%%%%%%%%%%%%%%%%%%%%%%%%
\begin{center}
\vspace*{2.5mm}
\begin{figure}[H]
%\centering
\hspace*{-0.45cm}
\subfloat{\includegraphics[scale=0.45]{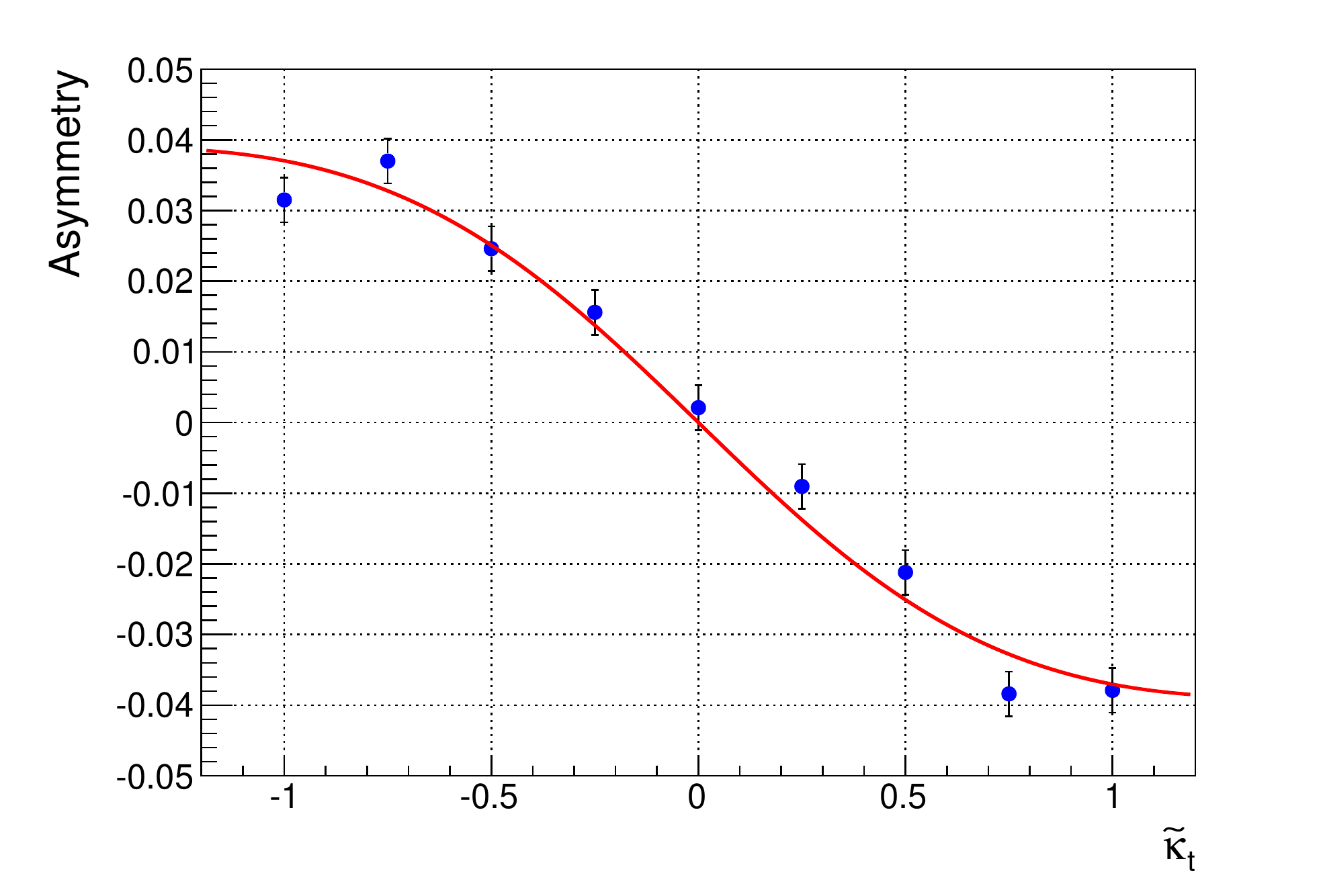}}
\hspace*{0.002\textwidth}
\subfloat{\includegraphics[scale=0.45]{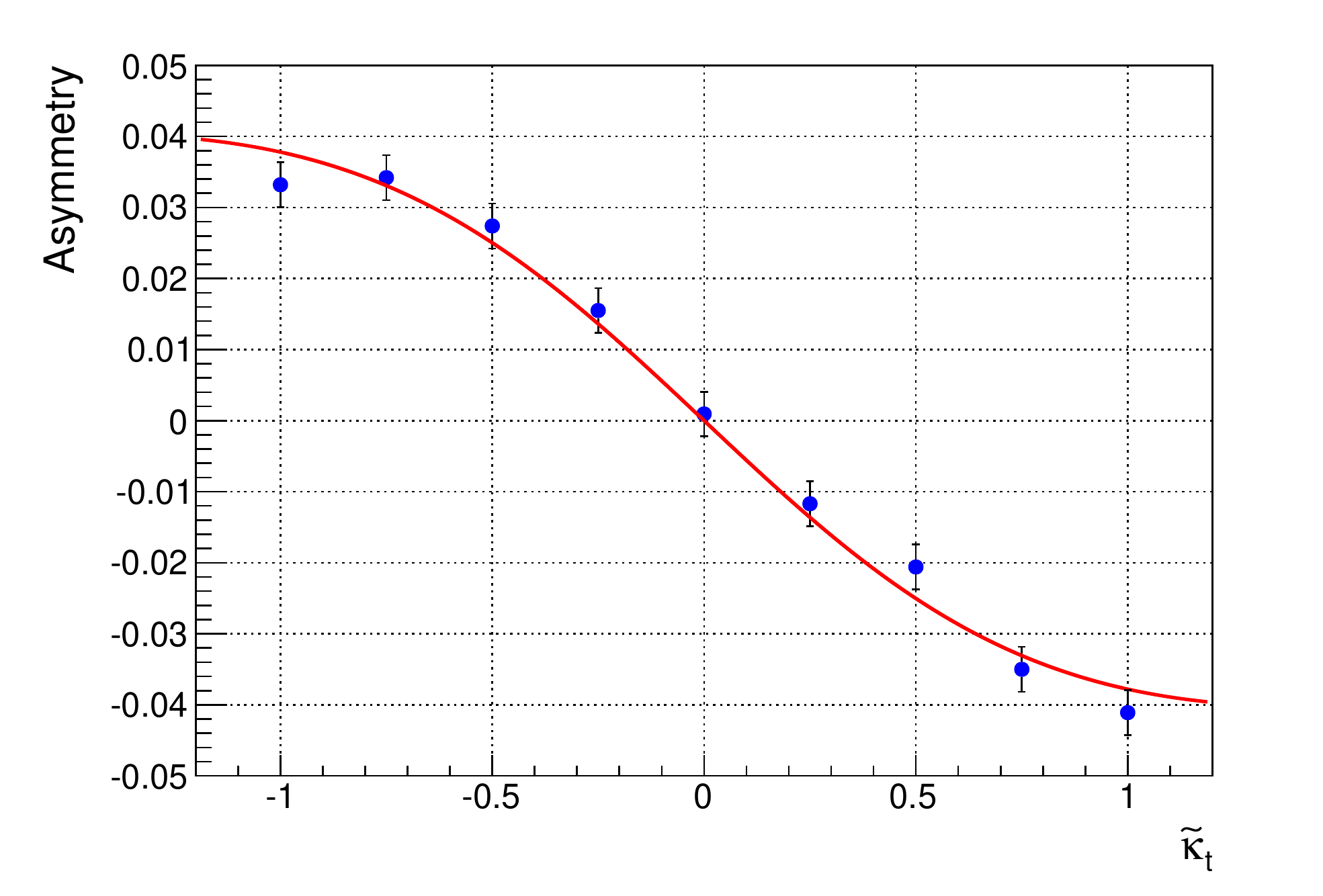}} \\
%\hspace*{0.03\textwidth}
\centering
\subfloat{\includegraphics[scale=0.45]{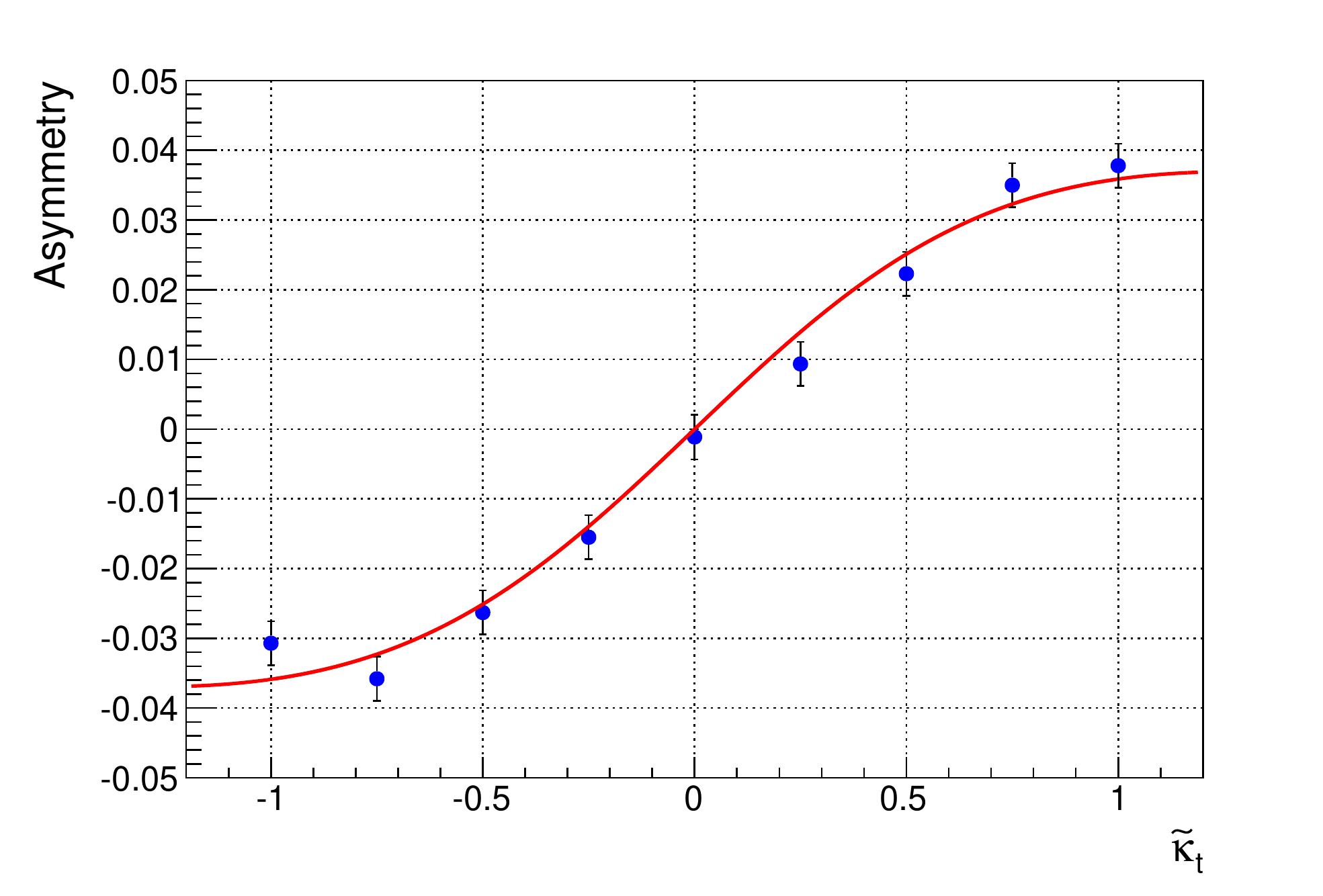}}
\caption{Asymmetries for the TPs
  $\epsilon_1=\epsilon(t,\tbar,n_t,n_{\tbar})$ (top-left),
  $\epsilon_2=\epsilon(Q,\tbar,n_t,n_{\tbar})$ (top-right) and
  $\epsilon_3=\epsilon(Q,t,n_t,n_{\tbar})$ (bottom). The points
  represent the values for $\kpt=0,\pm 0.25, \pm 0.5, \pm 0.75,\pm 1$
  and the red solid line is the fitting curve.}
\label{fig3}
\end{figure}
\end{center}
%%%%%%%%%%%%%%%%%%%%%%%%%%%%%%%%%%%%%%%%%%%%
The results shown in table~\ref{table1}
and figure~\ref{fig3} all assume a $pp$ initial state, which is
actually a combination of events coming from $gg$ and $q\qbar$
initial states.
While this combination of initial states is the appropriate scenario
to consider, it
is interesting to consider the relative contributions to the asymmetry
coming from the
$gg$ and $q\qbar$ initial states.  Figure~\ref{fig4} shows
three curves for the ``$\epsilon_1$'' case, one for $gg$-initiated
events, one for $q\qbar$-initiated events, and one for the usual
combination of these events (the ``$pp$'' initial state).
Interestingly, we see from figure~\ref{fig4} that the asymmetry for this TP
is enhanced for
$gg$-initiated production, while it is reduced and of opposite sign
for the $q\qbar$-initiated events. The asymmetry for the $pp$ case is
evidently
dominated by the $gg$ contribution, but is somewhat smaller
in magnitude due to the
$q\qbar$ contribution.
%%%%%%%%%%%%%%%%%%%%% FIGURA 4  %%%%%%%%%%%%%%%%%%%%%%%%%%%%
\begin{center}
\vspace*{-4mm}
\begin{figure}[H]
\centering
\includegraphics[scale=0.45]{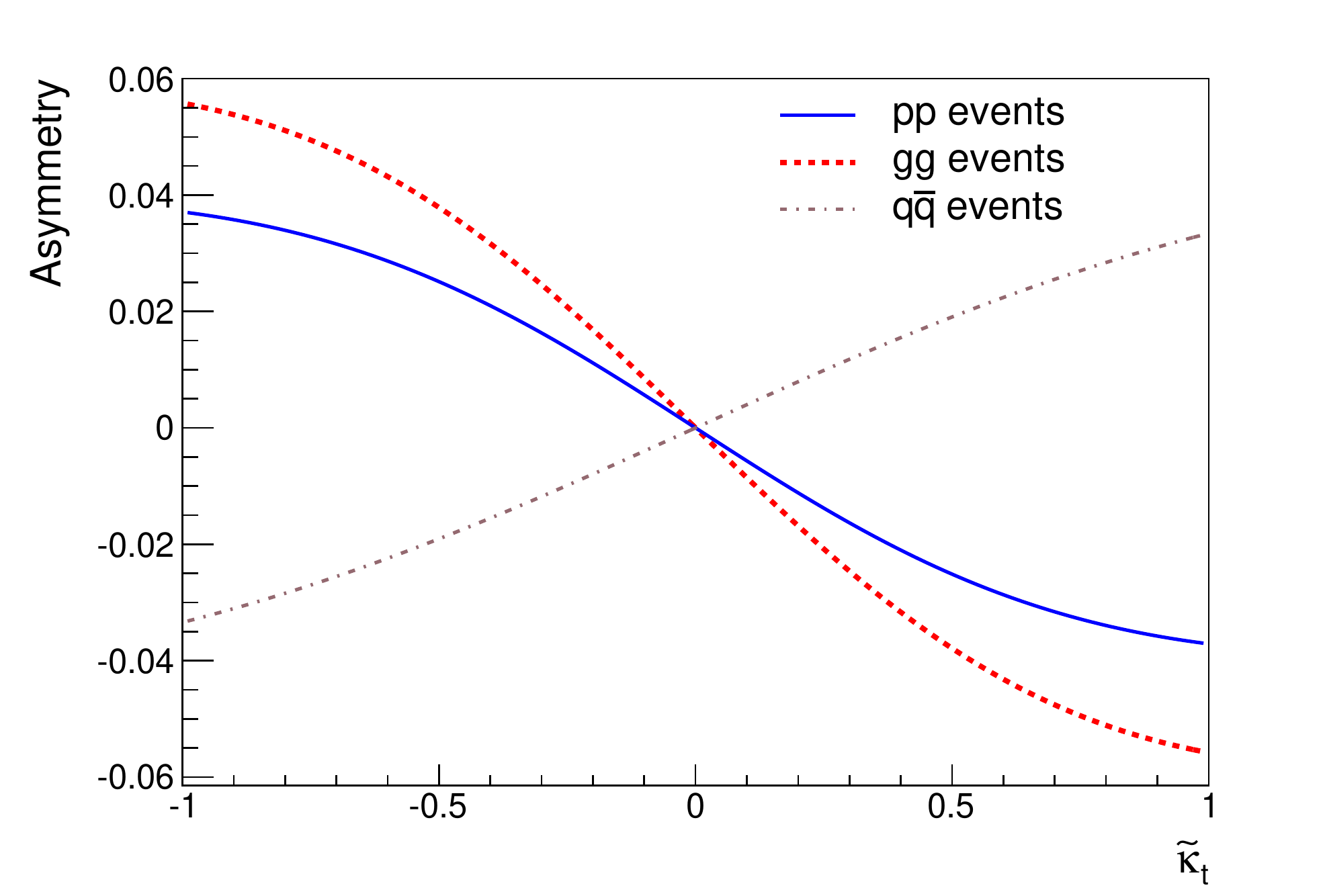}
\caption{Asymmetry for the TP
  $\epsilon_1=\epsilon(t,\tbar,n_t,n_{\tbar})$. The dashed line (red)
  corresponds to $gg$-initiated production, the dot-dashed line (grey)
  to $q\qbar$-initiated production and the solid line (blue) to $pp$
  production.}
\label{fig4}
\end{figure}
\end{center}
%%%%%%%%%%%%%%%%%%%%%%%%%%%%%%%%%%%%%%%%%%%%%%%%%%%%%%%%%%%%%
\vspace*{-6mm}

We have also tested various combinations of the TPs $\epsilon_{1,2,3}$ and have found
that the asymmetry is enhanced for the following combination:
\beq
\setlength{\abovedisplayskip}{9.5pt}
\setlength{\belowdisplayskip}{9.5pt}
\label{eq24}
\epsilon_4=\epsilon_3-\epsilon_2=\epsilon(Q,t-\tbar,n_t,n_{\tbar}).
\eeq
Note that in the $Q$ rest frame, $\epsilon_4=Q^0
(\vec{t}-\vec{\tbar}\,)\cdot(\vec{n}_t\times \vec{n}_{\tbar})$ and the
sign of this TP is determined by the quantity
$(\vec{t}-\vec{\tbar}\,)\cdot(\vec{n}_t\times \vec{n}_{\tbar})$.
The
values obtained for the asymmetry associated with this TP are shown in
table~\ref{table2}. By comparing the results in tables~\ref{table1}
and \ref{table2}, we see that the capability of this
asymmetry to distinguish between the
two $\mathrm{CP}$-mixed scenarios is increased by at least $2.8\sigma$.
\vspace{4mm}
\begin{table}[H]
\caption{Asymmetry for the TP $\epsilon_{4}$ for the SM case and the
  two $\mathrm{CP}$-mixed scenarios.  The
  values are obtained using sets of $10^5$ simulated events.
}
\label{table2}
\begin{center}
\begin{tabular}{|C{1cm}|C{1cm}||C{2cm}|C{2cm}|}
%\begin{tabular}{|c|r||r|c||r|c||r|c|}
\hhline{|====|}
%\hhline{|--------|}
$\kappa_t$&$\tilde{\kappa}_t$~~&$\mathcal{A}(\epsilon_4)$~~&$\mathcal{A}(\epsilon_4)/\sigma_{\mathcal{A}}$ \\ 
\hhline{|====|} 
%\hhline{|--------|}
$1$ & $-1$~~~ & $-0.0371$~~~ & $-11.7$~~ \\[0.6mm]
\hline
$1$ & $0$ & $0.0004$~ & $0.1$ \\[0.6mm]
\hline
$1$ & $1$ & $0.0461$~ & $14.6$ \\[0.6mm]
\hhline{|====|}
%\hhline{|--------|}
\end{tabular}
\end{center} 
\end{table}
\par Finally, it is worth noting that the asymmetries
described in this subsection are not useful for
discriminating between the SM hypothesis ($\kp=1,\kpt=0$) and the pure
pseudoscalar hypothesis ($\kp=0,\kpt=1$).  Since
the numerators of the 
asymmetries are linear in both
$\kp$ and $\kpt$, they are expected to vanish in these cases. However, we
will show in the next subsection
that there exist angular distributions
derived from the TPs that are actually suitable for distinguishing between these
two hyphotheses.
%%%%%%%%%%%%%%%%%%%%%%%%%%%%%%%%%%%%%%%%%%%%%%%%%
\subsection{Angular Distributions}
\label{sec3.2}

Given a certain TP, it is possible to define associated angular
distributions that are sensitive to the pseudoscalar coupling
$\kpt$. In order to clarify this, let us first consider the TP
$\epsilon(t,\tbar,n_t,n_{\tbar})$. This TP can be written as
$\epsilon(t+\tbar,\tbar,n_t,n_{\tbar})$, so that in the reference frame
defined by $\vec{t}+\vec{\tbar} =0$ and $\vec{\tbar}\parallel \hat{z}$
we have
\beq
\label{eq25}
\epsilon(t+\tbar,\tbar,n_t,n_{\tbar})=M_{t\tbar}\,|\vec{\tbar}|\,(\vec{n}_t\times \vec{n}_{\tbar})_z=M_{t\tbar}\,|\vec{\tbar}||\vec{n}_t||\vec{n}_{\tbar}|\sin\theta_{n_t}\sin\theta_{n_{\tbar}}\sin \Delta \phi(n_t,n_{\tbar}),
\eeq
where $M_{t\tbar}$ is the invariant mass of the $t\tbar$ pair, the
angles $\theta_{n_t}$ and $\theta_{n_{\tbar}}$ denote the polar angles
of $\vec{n}_t$ and $\vec{n}_{\tbar}$, respectively, and 
$\Delta\phi(n_t,n_{\tbar})$ is the angular difference between the
projections of $\vec{n}_t$ and $\vec{n}_{\tbar}$ onto the plane
perpendicular to $\vec{\bar{t}}$. If we define the angle
$\Delta\phi(n_t,n_{\tbar})$  to be within the
range $[-\pi,\pi]$, we see from eq.~(\ref{eq25}) that its sign will
determine the sign of the TP. Thus, the distribution of the
number of events with respect to the angle
$\Delta\phi(n_t,n_{\tbar})$ is related to the asymmetry of the TP,
\beq
\label{eq26}
\mathcal{A}(\epsilon)=1-2\frac{N(\epsilon < 0)}{N_T}\,\,\mbox{ and }\,\,\frac{N(\epsilon < 0)}{N_T}=\int^0_{-\pi}\frac{1}{N_T}\frac{dN}{d\Delta\phi(n_t,n_{\tbar})}\,d\Delta\phi(n_t,n_{\tbar}),
\eeq  
where $N_T$ is the total number of events.
Moreover, for a certain TP
one can derive different angular distributions by considering
different reference frames, although all of these will satisfy
eq.~(\ref{eq26}) (note that $\mathcal{A}(\epsilon)$
is Lorentz invariant). Recalling the various TPs considered
in section~\ref{sec2}, we examine the following angular distributions.
\begin{enumerate}
\item {\boldmath $\epsilon_1 = \TPa$.}  To probe $\epsilon_1$, we construct
  the distribution
  $d\sigma/d\Delta\phi_1(n_t,n_{\tbar})$ in the rest frame of $t\tbar$,
  taking $\vec{\tbar}$ to define the $z$-axis. The angle
  $\Delta\phi_1(n_t,n_{\tbar})$ is the angular difference between the
  projection of the spin vectors onto the plane perpendicular to
  $\vec{\tbar}$.
\item {\boldmath $\epsilon_2 = \TPb$.}  In this case, we
  define the distribution
  $d\sigma/d\Delta\phi_2(n_t,n_{\tbar})$ in the rest frame of $Q$, taking
  $\vec{\tbar}$ to define the $z$-axis. The angle
  $\Delta\phi_2(n_t,n_{\tbar})$ is the angular difference between the
  projection of the spin vectors onto the plane perpendicular to
  $\vec{\tbar}$.
\item {\boldmath $\epsilon_3 = \TPc$.}  The distribution
  $d\sigma/d\Delta\phi_3(n_t,n_{\tbar})$ is also defined
  in the rest frame of $Q$, but this time taking
  $\vec{t}$ to be along the $z$-axis. The angle $\Delta\phi_3(n_t,n_{\tbar})$
  is the angular difference between the projection of the spin vectors
  onto the plane perpendicular to $\vec{t}$.
\end{enumerate}
\par
Figure~\ref{fig5} shows the normalized distributions obtained
for the first case listed above.
Four scenarios are considered, corresponding to the SM
($\kp= 1$ and $\kpt=0$), two cases in which the Higgs boson
has mixed $\mathrm{CP}$ couplings ($\kp= 1$ and $\kpt=\pm 1$)
and a case in which the Higgs boson is purely 
$\mathrm{CP}$-odd ($\kp= 0,\kpt=1$). Figure~\ref{fig6} shows
the analogous distributions for $\epsilon_2$. The distributions
corresponding to $\epsilon_3$ are similar to those of $\epsilon_2$,
except that the ``shifts'' are in the opposite directions for
the two $\mathrm{CP}$-mixed cases.  Given the similarities of the
plots we do not include them here. 
\par As can be seen from figures~\ref{fig5} and \ref{fig6}, the
peaks of the 
distributions are shifted to the left or the right of the origin
in the $\mathrm{CP}$-mixed
cases ($\kp =1$ and $\kpt=\pm 1$).
The magnitude of the shift appears to be approximately the
same in both cases, but is in the
opposite direction for $\kp=\kpt=1$ compared to $\kp=-\kpt=1$, thus allowing
one to distinguish
the sign of the pseudoscalar coupling. The observed dependence
on the sign of $\kpt$ in these cases
is consistent with the fact that the numerator of $\mathcal{A}(\epsilon)$
is linear in $\kpt$ (see eq.~(\ref{eq23}))
and that the quantity $N(\epsilon < 0)/N_T$ is related
to the angular distribution according to eq.~(\ref{eq26}). The angular
distributions for the SM case ($\kp =1$ and $\kpt=0$) and
the pure pseudoscalar case ($\kp =0$ and $\kpt= 1$) are
visibly different from each other and from the $\mathrm{CP}$-mixed
scenarios. Comparing the SM and purely pseudoscalar cases,
we note that while the angular distributions for the former case
exhibit a minimum at $\Delta\phi_{1,2}(n_t,n_{\tbar})=0$, those
for the latter case exhibit a peak at this location.
Thus, these two scenarios can be distinguished from each other
via these angular distributions.  This is to be contrasted
with the situation for the asymmetries $\mathcal{A}(\epsilon)$, which vanish
in both cases. \par
In order to quantify the shifts discussed above, we have fitted
the simulated distributions with the following function, which was proposed in
ref.~\cite{Ellis},
\beq
\label{eq27}
\frac{1}{\sigma}\frac{d\sigma}{d\Delta\phi_i(n_t,n_{\tbar})}=a_0 + a_1\cos(\Delta\phi_i(n_t,n_{\tbar})+\delta),\qquad\, i=1,2,3.
\eeq
%
%%%%%%%%%%%%%%%%%%%%%%%%% FIGURA 5 %%%%%%%%%%%%%%%%%%%%%%%%%
%%%% ruta turin facultad: /home/nico/Documentos/ttbarH/Material_final/ todas las figs para abajo
\begin{center}
\vspace*{1.4mm}
\begin{figure}[H]
%\centering
\hspace*{-0.55cm}
%\subfloat{\includegraphics[scale=0.443]{TP1_10_nuevo.pdf}}
\subfloat{\includegraphics[scale=0.45]{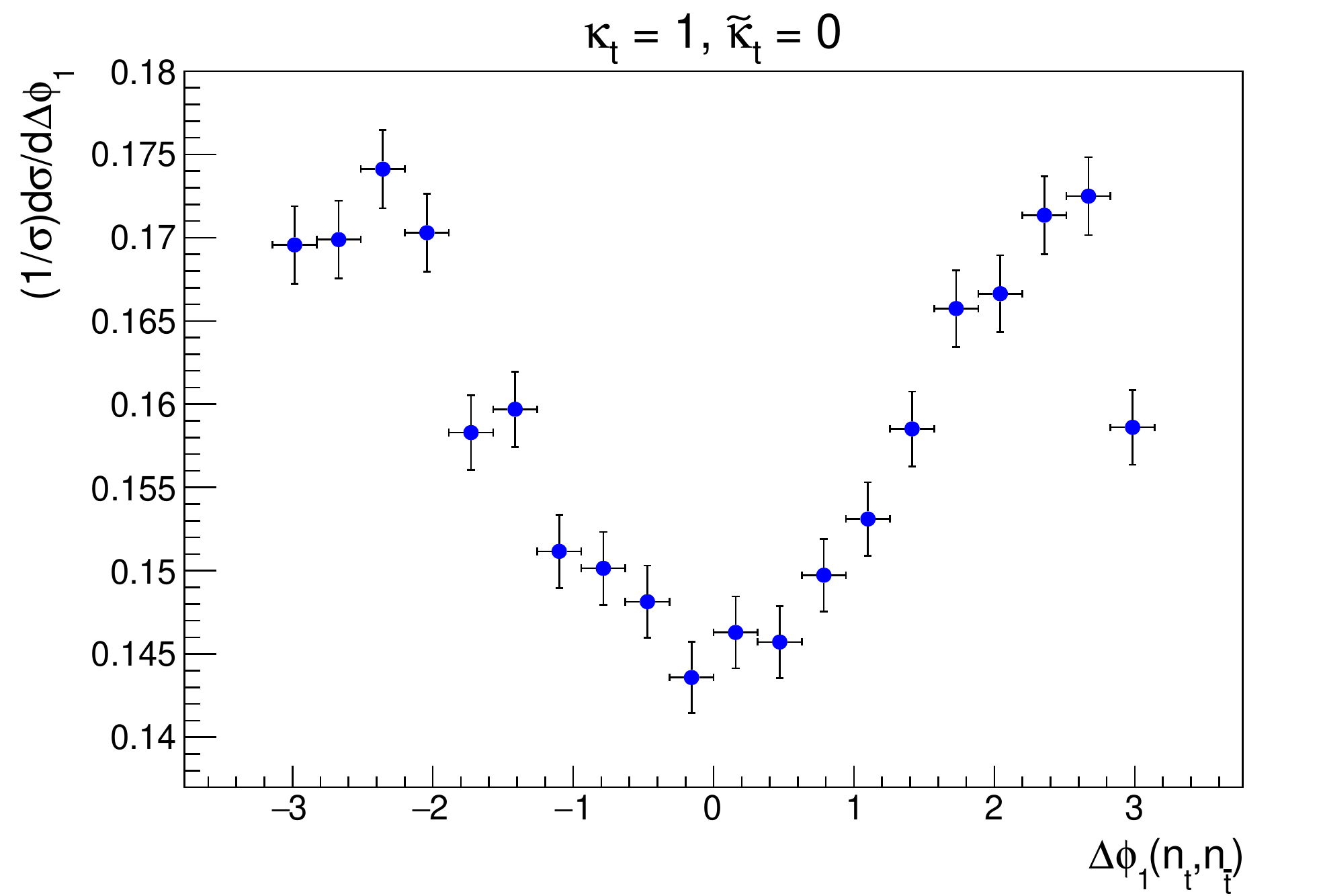}}
\hspace*{-0.006\textwidth}
\subfloat{\includegraphics[scale=0.45]{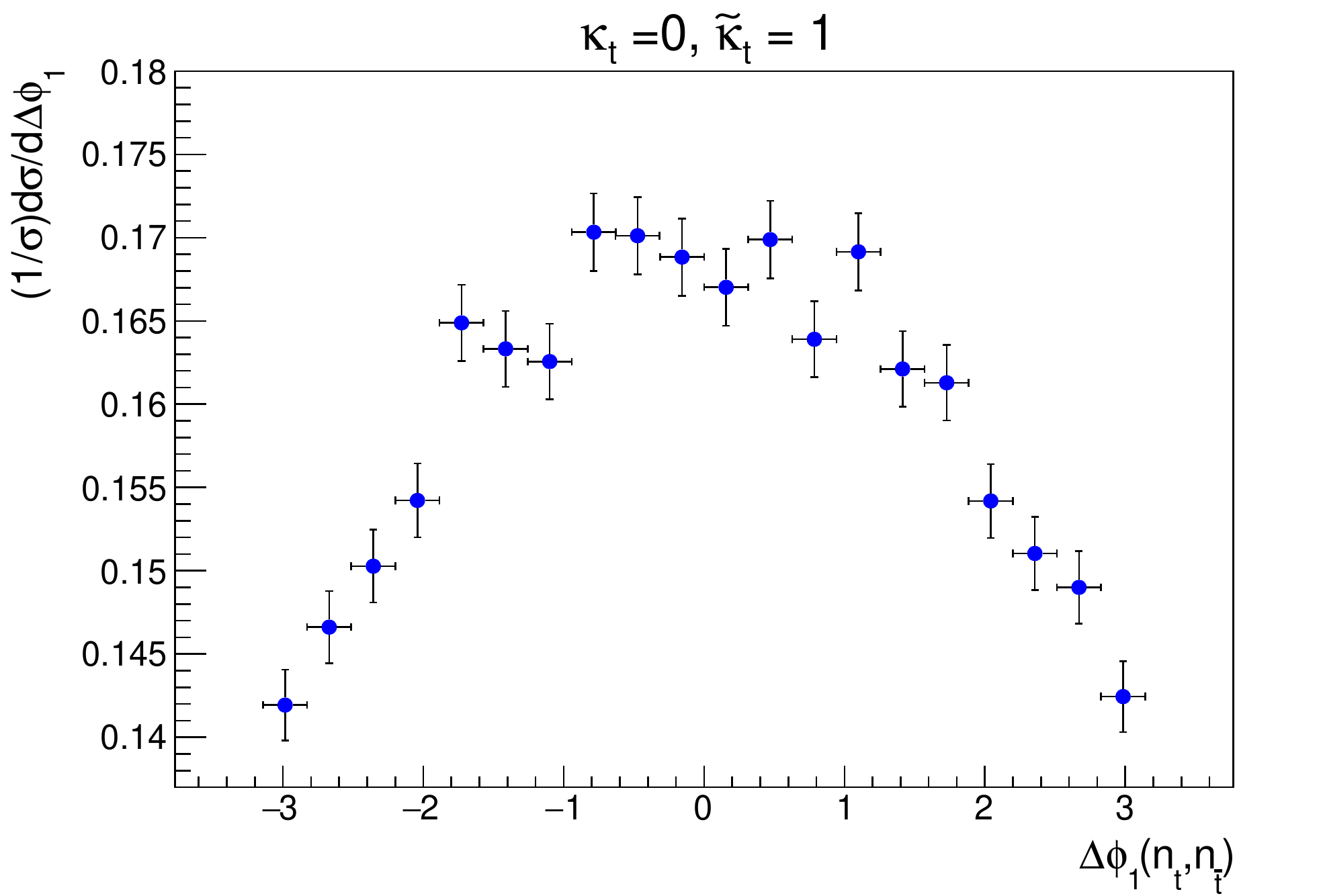}} \\
\hspace*{-0.55cm}
%\hspace*{0.03\textwidth}
\subfloat{\includegraphics[scale=0.45]{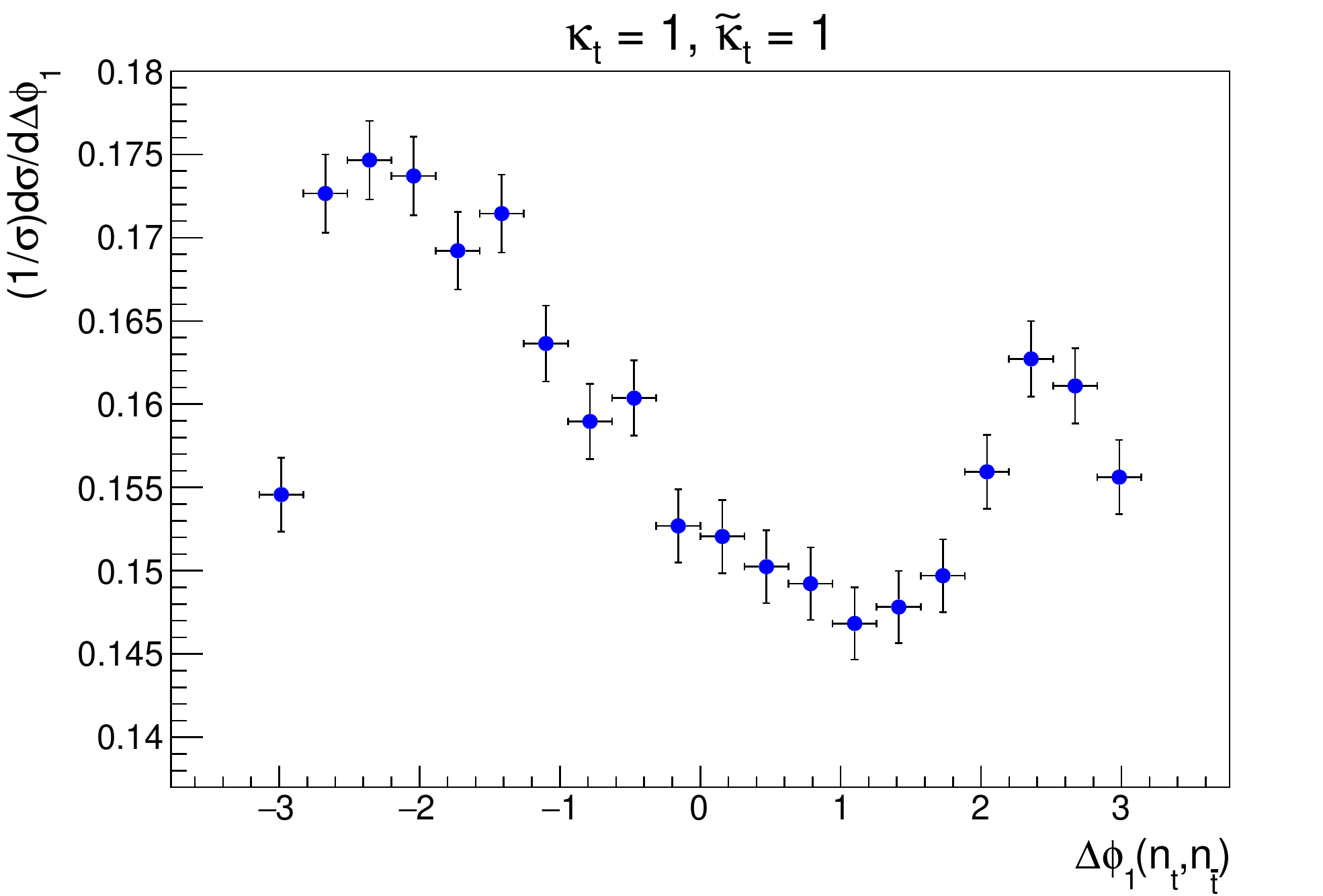}}
\hspace*{-0.006\textwidth}
\subfloat{\includegraphics[scale=0.45]{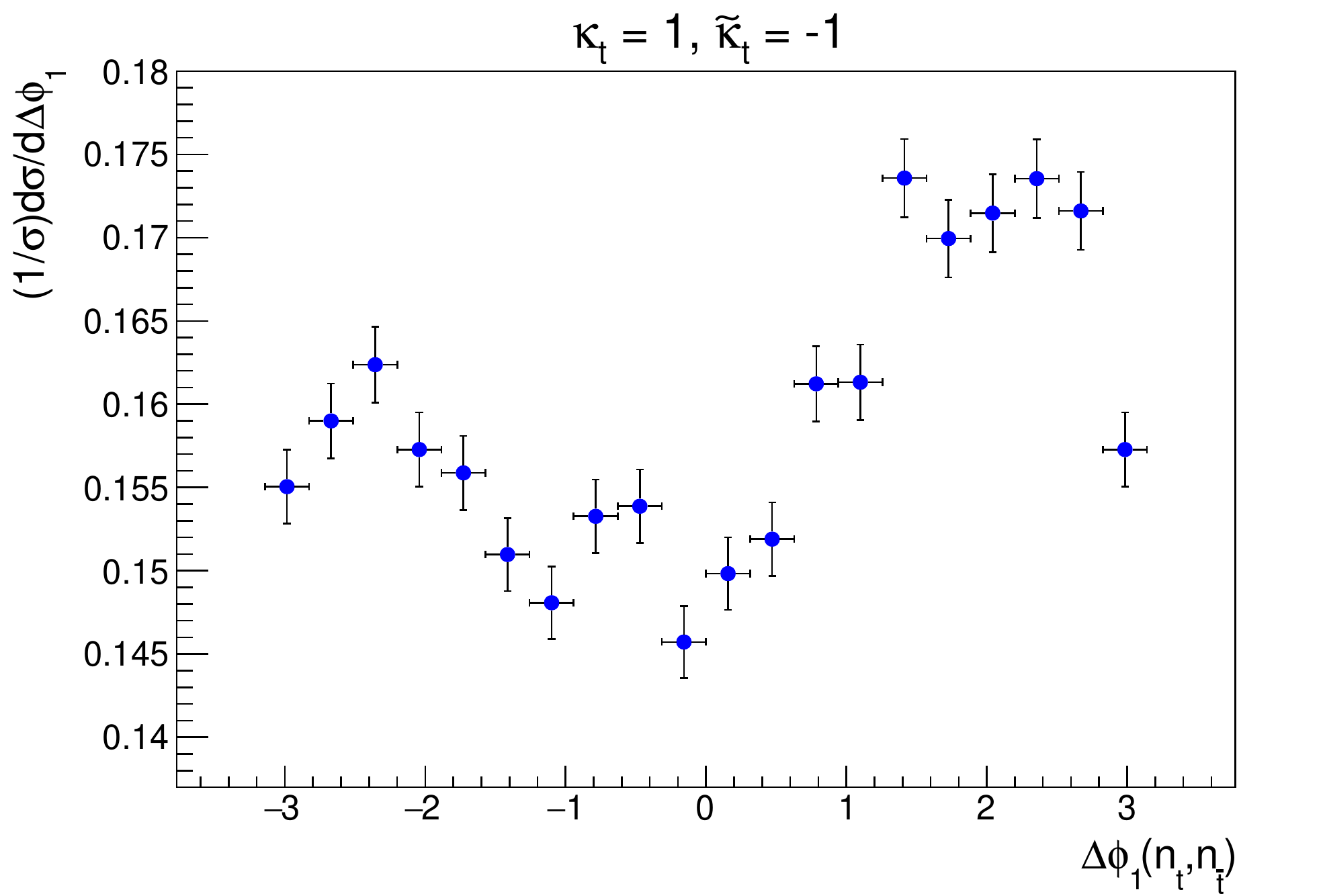}}
\vspace*{1cm}
\caption{Angular distributions associated with the TP
  $\epsilon_1 = \epsilon(t,\tbar,n_t,n_{\tbar})$ for various values of
  $\kp$ and $\kpt$. The error bars correspond to the statistical
  uncertainties.}
\label{fig5}
\end{figure}
\end{center}
%%%%%%%%%%%%%%%%%%%%%%%%% FIGURA 6 %%%%%%%%%%%%%%%%%%%%%%%%%
\begin{center}
%\vspace*{-4mm}
\begin{figure}[H]
%\centering
\hspace*{-0.52cm}
\subfloat{\includegraphics[scale=0.45]{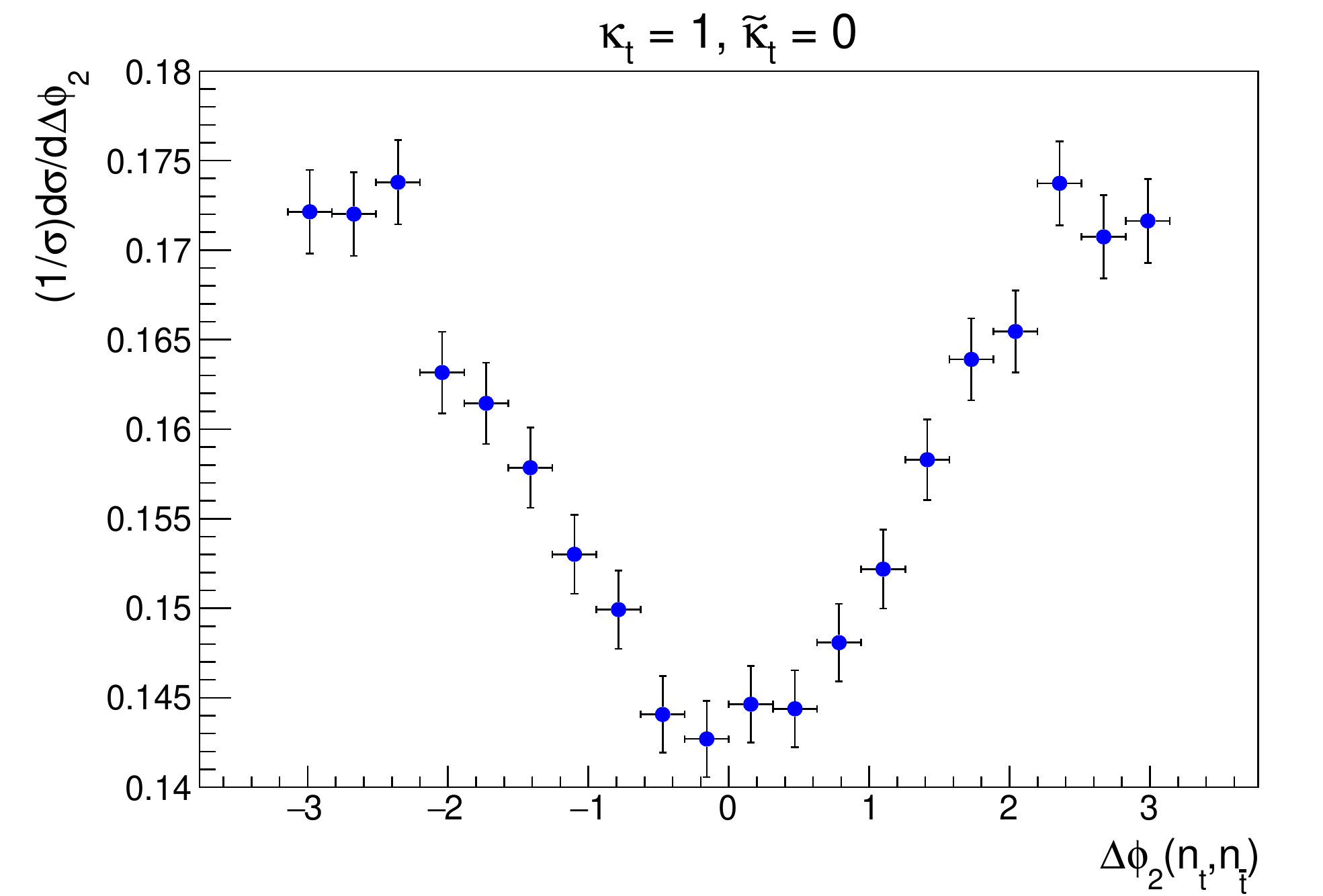}}
\hspace*{-0.006\textwidth}
\subfloat{\includegraphics[scale=0.45]{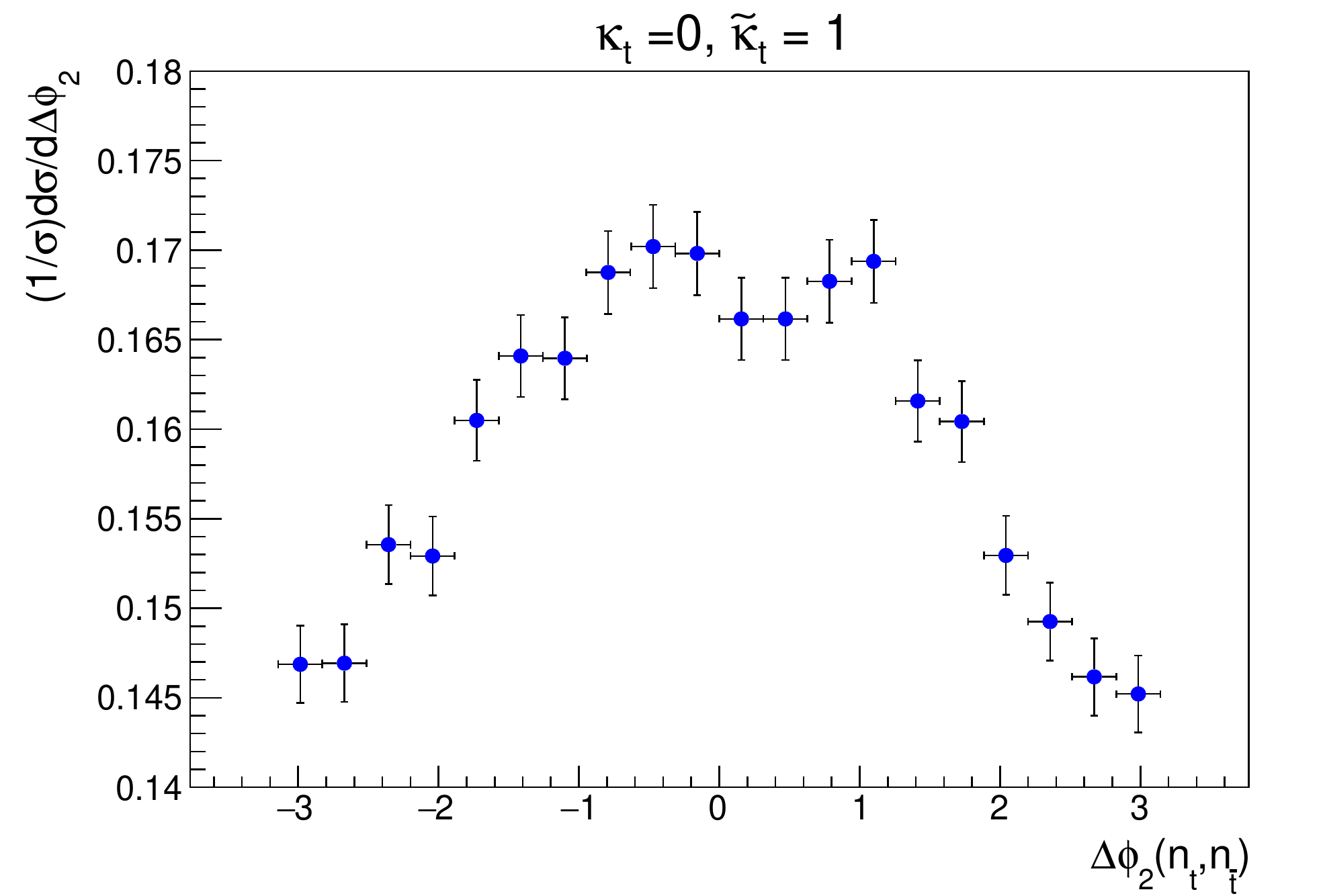}} \\
\hspace*{-0.56cm}
%\hspace*{0.03\textwidth}
\subfloat{\includegraphics[scale=0.45]{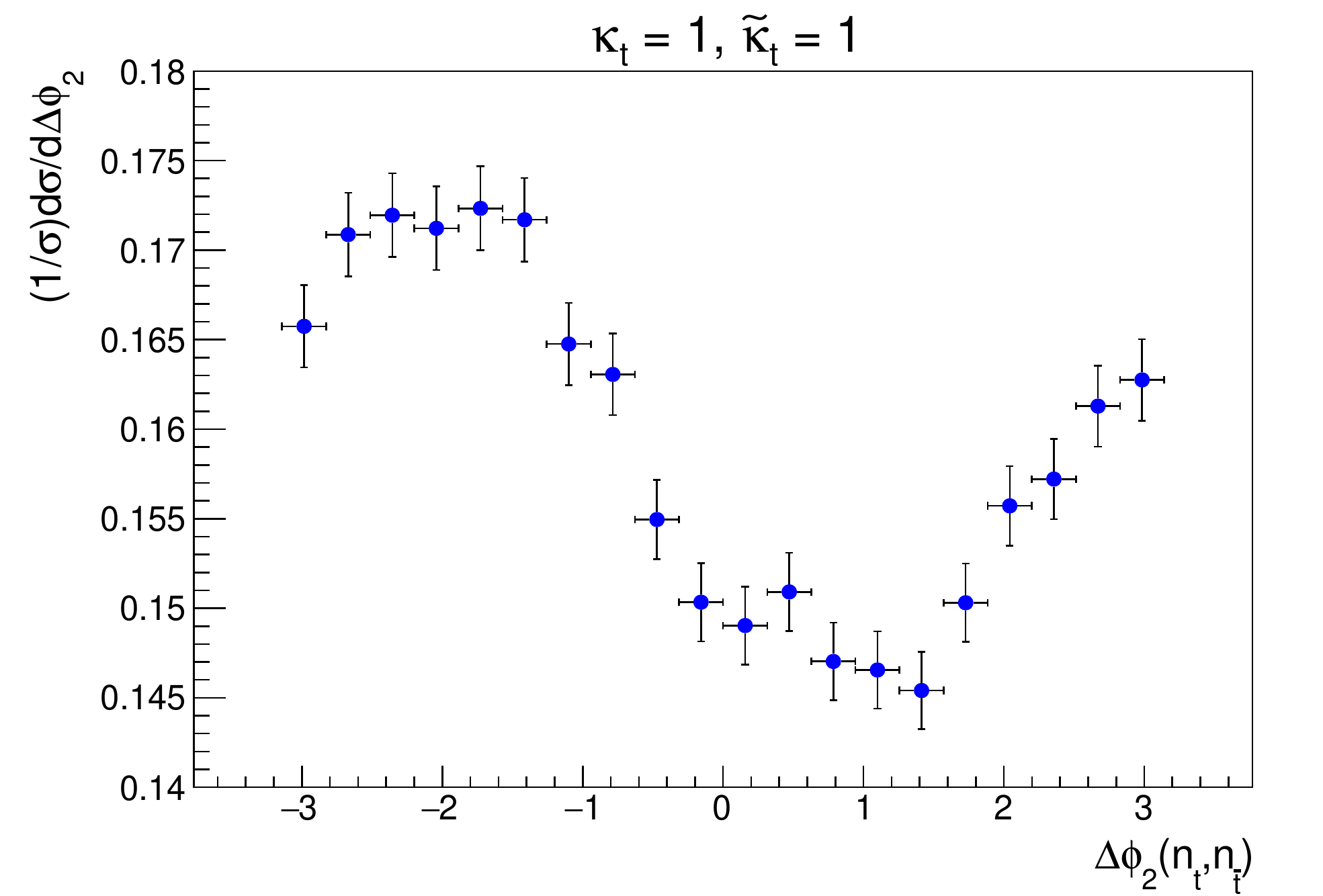}}
\hspace*{0.002\textwidth}
\subfloat{\includegraphics[scale=0.45]{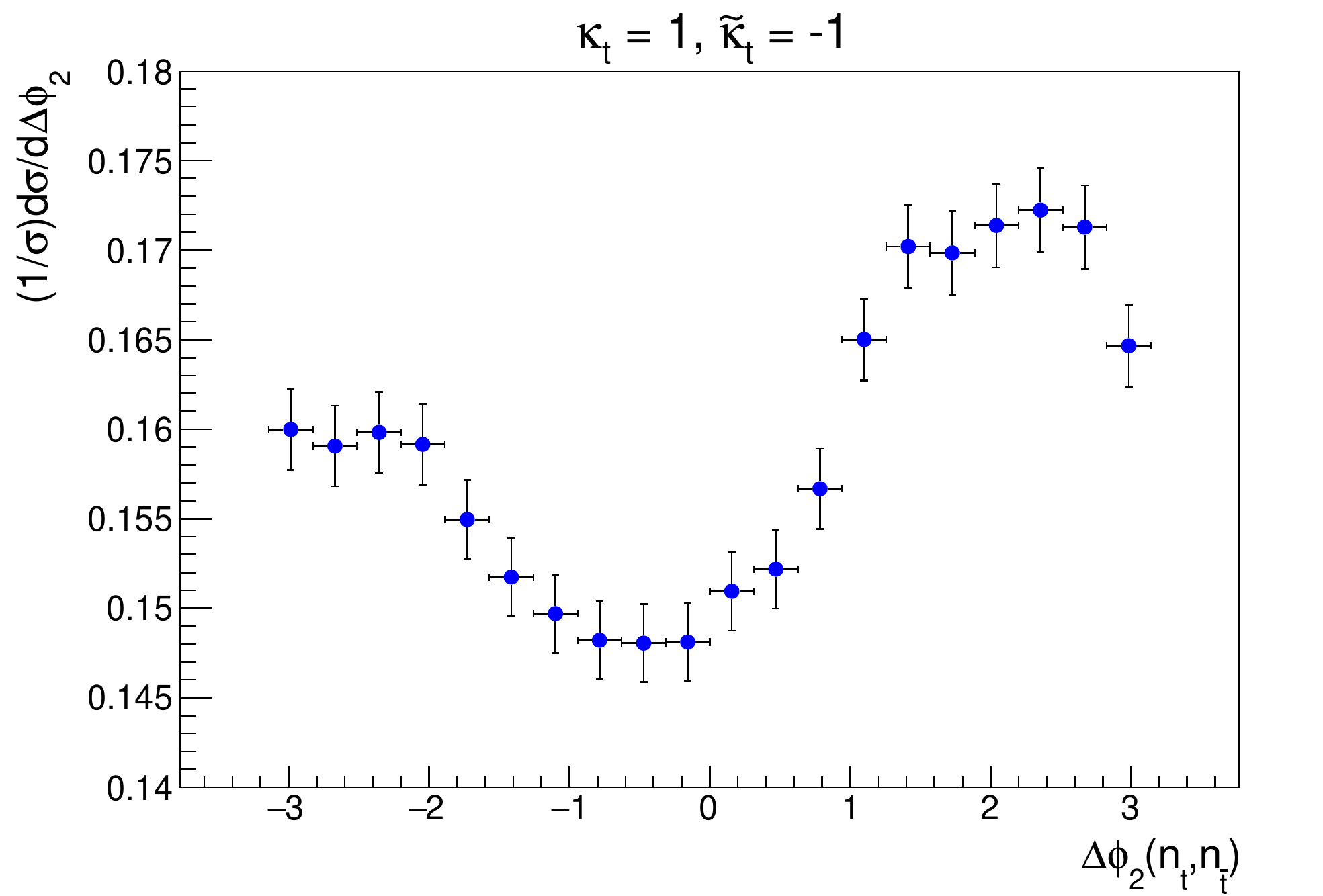}}
%\subfloat{\includegraphics[scale=0.45]{TP2_1-1_nuevo.pdf}}
\caption{Angular distributions associated with the TP $\epsilon_2 = \TPb$ for
  various values of $\kp$ and $\kpt$. The error bars indicate the statistical
  uncertainties.}
\label{fig6}
\end{figure}
\end{center}
To the extent that the above expression is exact, we note that
eq.~(\ref{eq26}) 
gives $\mathcal{A}(\epsilon_i)=-4a_1 \sin\delta$. With
this fitting function, we obtain phase shifts $\delta$ that
are approximately between
$0.9$ and $1$ ($-1$ and $-0.9$) for $\kp=-\kpt=1$ ($\kp=\kpt=1$), both
for $\epsilon_1$ and $\epsilon_2$.\footnote{The results for the
  TP $\epsilon_3$ are relatively similar to
  those for $\epsilon_2$, except that the phase shifts have the opposite
  sign in the $\mathrm{CP}$-mixed cases.  Given this similarity
  we do not include the corresponding results for the $\epsilon_3$ distribution
  here.}  However, the quality of the fits in the four scenarios considered is
not very good, particularly for $\epsilon_1$.  The
$\chi^2/\mathrm{d.o.f}$ for the fits corresponding to $\epsilon_1$ are in the range
$1.69$-$3.86$, while for $\epsilon_2$ they are in the range $0.53$-$1.16$.
The deviation from the functional
form proposed in eq.~(\ref{eq27}) appears to be due primarily to the $\Delta
R_{ll}$ cut that we have imposed. In fact, when this cut is turned off, the
above ranges for the $\chi^2/\mathrm{d.o.f}$ become $0.75$-$1.14$ and
$0.44$-$1.07$ for the $\epsilon_1$ and $\epsilon_2$ distributions,
respectively. tables \ref{table3} and \ref{table4} list the
results of the fits obtained when the $\Delta R_{\ell\ell}$ cut is
relaxed.  Figure~\ref{fig7} shows the corresponding
plots for a couple of the scenarios.
As is evident from tables \ref{table3} and \ref{table4}, the
parameter $\delta$ is sensitive not only to the modulus of $\kpt$ but
also to its sign, as would be expected from eq.~(\ref{eq26}).
The phase shift $\delta$ for the $\Delta\phi_1$ distribution
appears to exhibit a slightly higher sensitivity than
that obtained for the $\Delta\phi_2$ distribution, although the corresponding
numerical values obtained for the various scenarios are
compatible to within their statistical uncertainties. It is
important to stress, however, that the fits for the $\Delta\phi_2$ distributions
always yield smaller values for the $\chi^2/\mathrm{d.o.f}$.\par
In section~\ref{sec3.1} we defined a fourth triple product,
$\epsilon_4 = \epsilon_3-\epsilon_2$. We have constructed
an angular distribution related to this TP as well.
Specifically, we have analyzed the
$\Delta\phi(n_t,n_{\tbar})$ distribution in the $Q$ rest frame,
taking
$H$ to define the $z$-axis.  We have studied the distributions
for various values of $\kp$ and $\kpt$ and have
found that they are not well described by eq.~(\ref{eq27}).
Instead of resembling sinusoids that are shifted to the left or
right for different values of the parameters, the distributions
become distorted in such a way that there is a non-zero
asymmetry (see eq.~(\ref{eq26})).  Moreover, the associated asymmetry values are
larger than the asymmetries for the other TPs (see 
tables \ref{table1} and \ref{table2}). 
%%%%%%%%%%%%%%%%%%%%%%%%% FIGURA 7 %%%%%%%%%%%%%%%%%%%%%%%%%
\begin{center}
\vspace*{-2.4cm}
\begin{figure}[H]
%\centering
\hspace*{-0.52cm}
\subfloat{\includegraphics[scale=0.45]{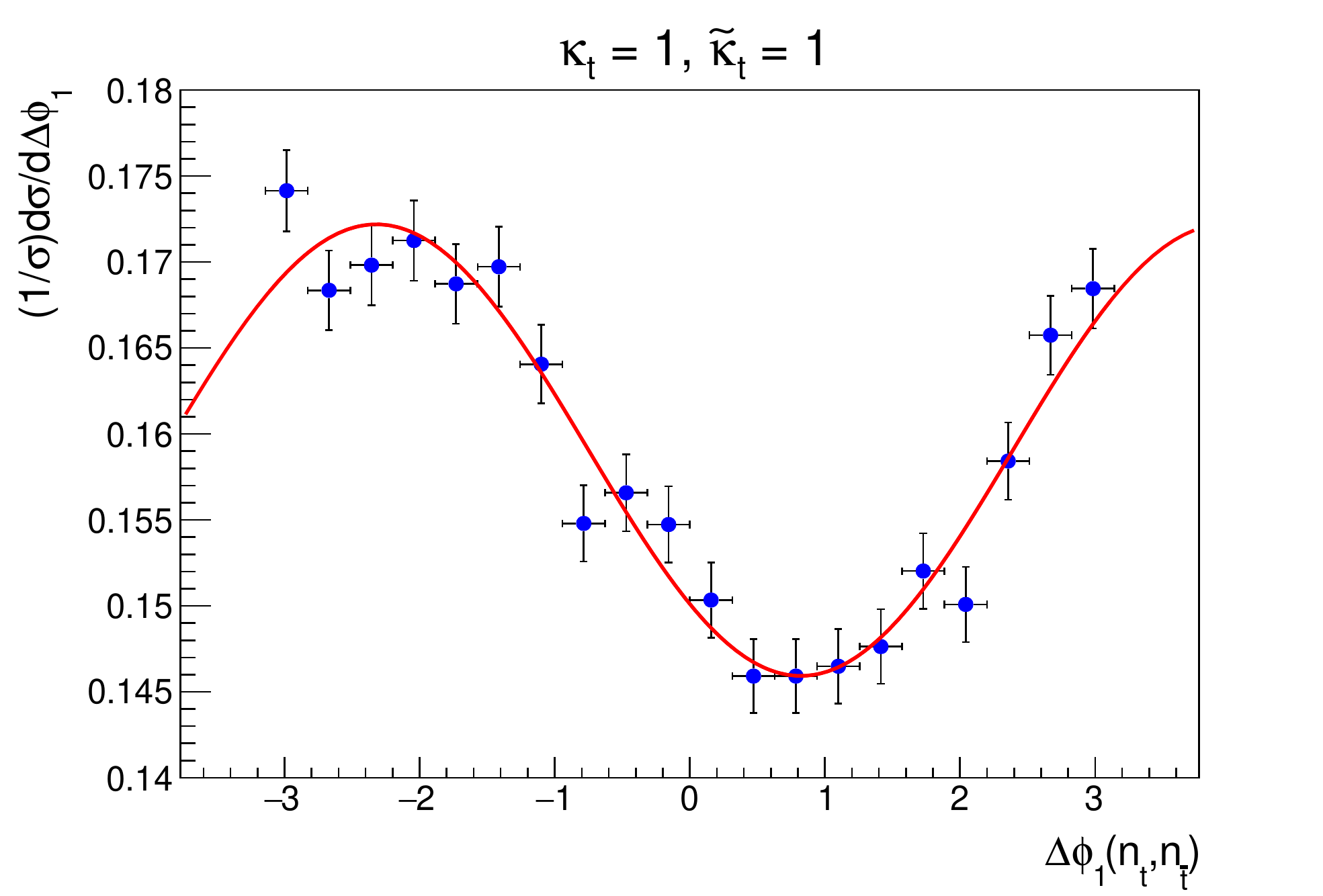}}
\hspace*{-0.006\textwidth}
\subfloat{\includegraphics[scale=0.45]{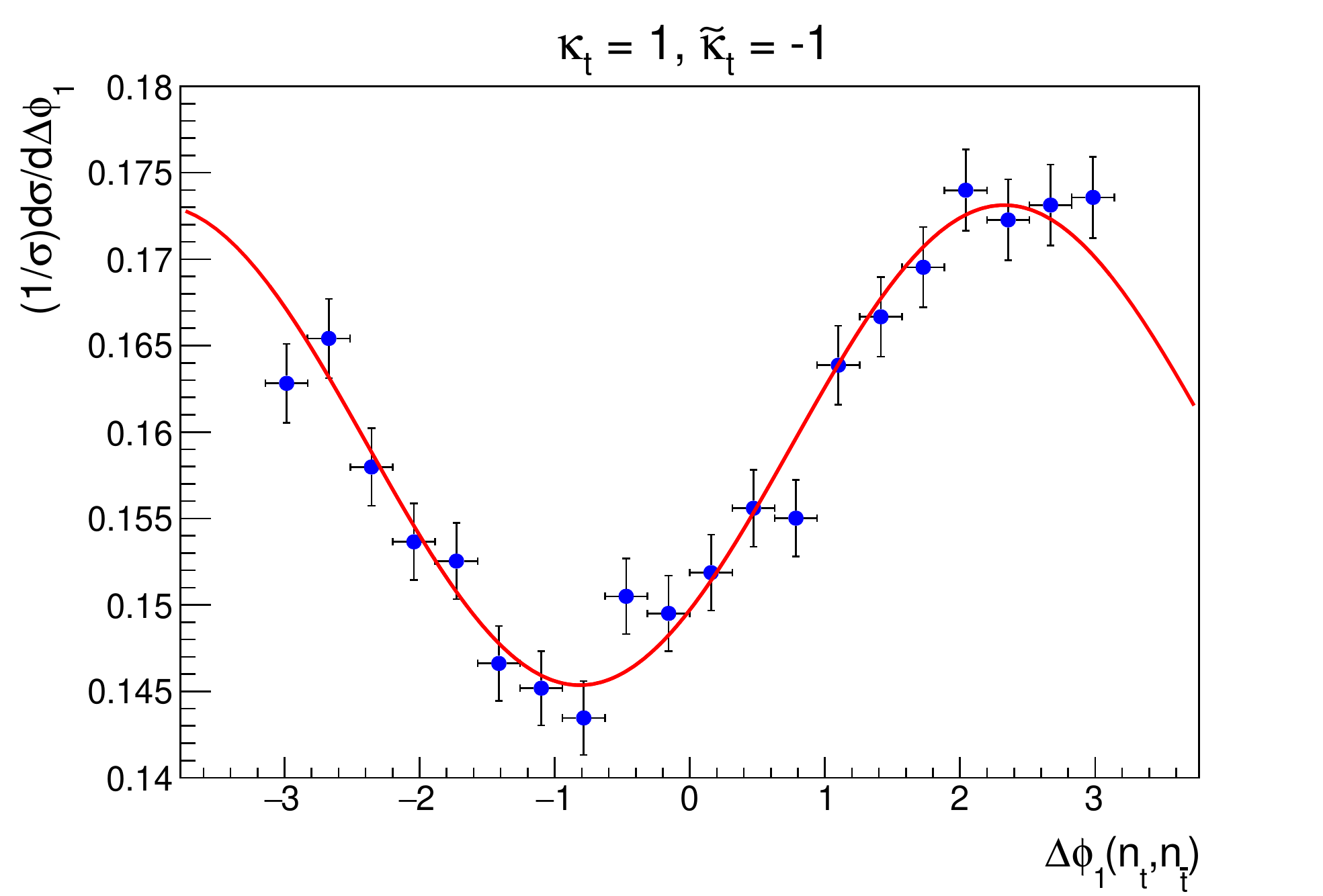}} \\
\hspace*{-0.52cm}
%\hspace*{0.03\textwidth}
\subfloat{\includegraphics[scale=0.45]{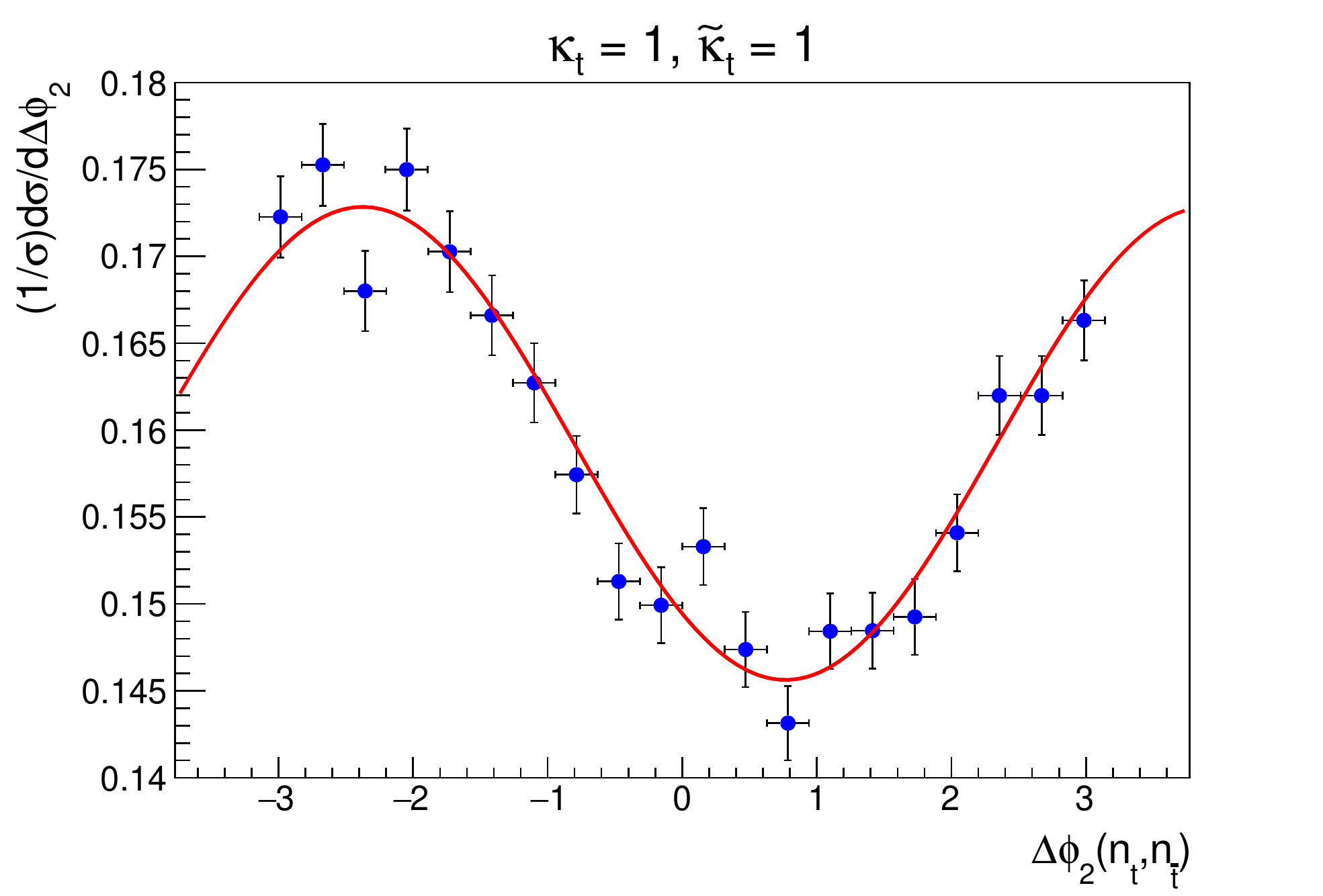}}
\hspace*{-0.006\textwidth}
\subfloat{\includegraphics[scale=0.45]{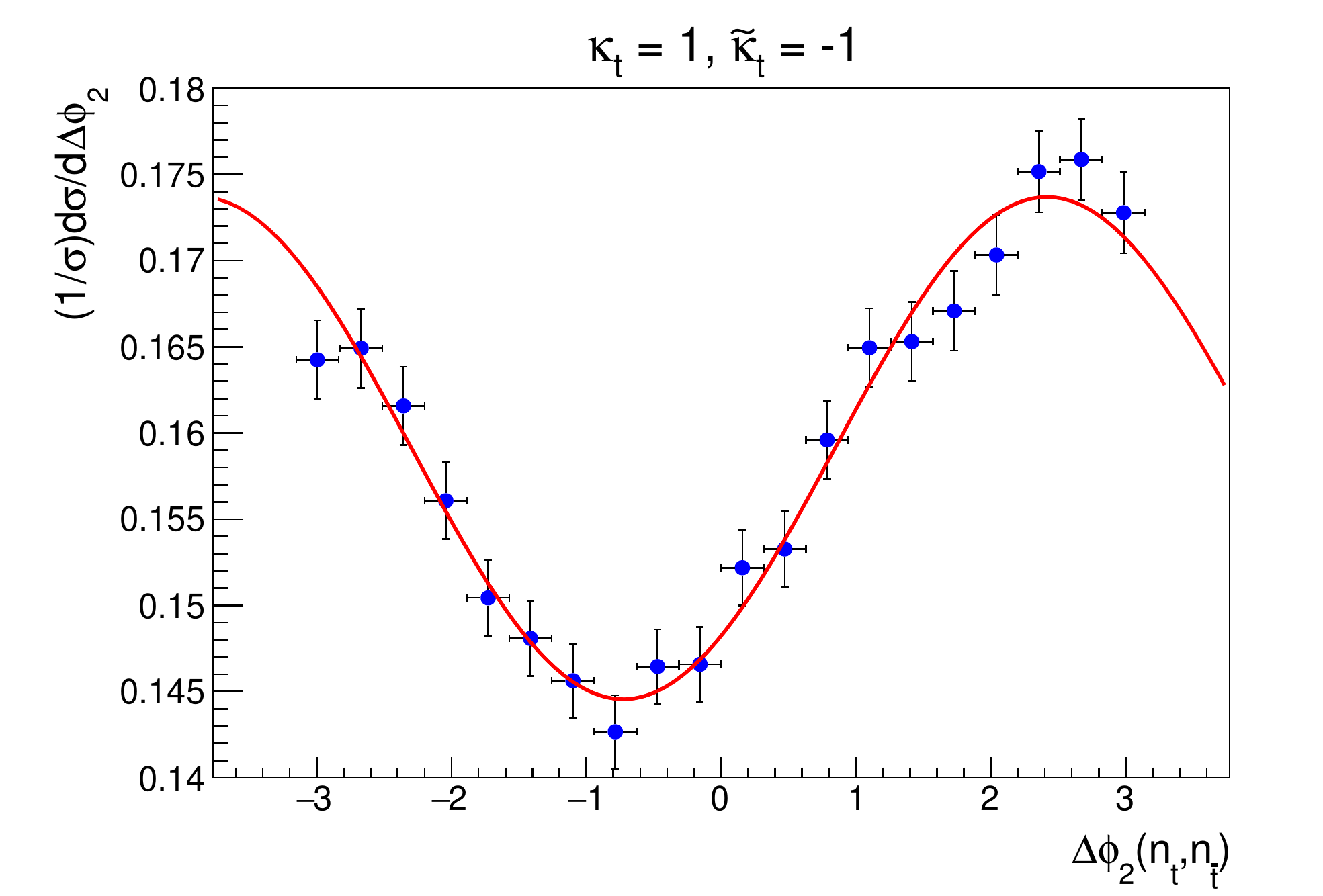}}
\caption{Angular distributions $d\sigma/(\sigma
  d\Delta\phi_1(n_t,n_{\tbar}))$ (top) and $d\sigma/(\sigma
  d\Delta\phi_2(n_t,n_{\tbar}))$ (bottom) associated with the TPs
  $\epsilon_1=\TPa$ and $\epsilon_2=\TPb$, respectively, for the
  $\mathrm{CP}$-mixed cases $\kp=\kpt=1$ (left) and $\kp=-\kpt=1$
  (right).  The $\Delta R_{\ell\ell}$ cut was turned off when
  generating these results. The
  corresponding fit curves (see eq.~(\ref{eq27})) are displayed in red.}
\label{fig7}
\end{figure}
\end{center}
\renewcommand{\arraystretch}{1.2}
\begin{table}[H]
\caption{Fit results for the angular distribution $d\sigma/(\sigma
  d\Delta\phi_1(n_t,n_{\tbar}))$ (related to the TP
  $\epsilon_1=\epsilon(t,\tbar,n_t,n_{\tbar})$) with the $\Delta
  R_{\ell\ell}$ cut turned off. Note that the sign of the parameter $a_1$
  changes for $\kp=0,\kp=1$, compared to the
  other cases.  We restrict $\delta$ to be between
    $\pm \pi/2$.}
\label{table3}
\begin{center}
\begin{tabular}{|C{1cm}|C{1cm}||C{3cm}|C{3cm}|C{3cm}|}
%\begin{tabular}{|c|r||r|c||r|c||r|c|}
\hhline{|=====|}
%\hhline{|--------|}
$\kappa_t$&$\tilde{\kappa}_t$~~&$a_0$~~&$a_1$& $\delta$~~ \\ 
\hhline{|=====|} 
%\hhline{|--------|}
$1$ & $-1$~~~ & $0.1592 \pm 0.0006$ & $-0.0139 \pm 0.0008$ & $0.81 \pm 0.07$ \\[0.6mm]
\hline
$1$ & $0$ & $0.1595 \pm 0.0006$ & $-0.0181 \pm 0.0008$ & $0.002 \pm 0.06\,\,$ \\[0.6mm]
\hline
$1$ & $1$ & $0.1591 \pm 0.0006$ & $-0.0131 \pm 0.0008 $ & $\,-0.82 \pm 0.07\quad$  \\[0.6mm]
\hline
$0$ & $1$ & $0.1591 \pm 0.0006$ & ~~$\,0.0102 \pm 0.0008$ & $0.11 \pm 0.08$ \\
\hhline{|=====|}
%\hhline{|--------|}
\end{tabular}
\end{center} 
\end{table}
%%%%%%%%%%%%%%%%%%%%%%%%%%%%%%%%%%%%%%%%%%%%%%%%%%%%%
\begin{table}[H]
\caption{Fit results for the angular distribution $d\sigma/(\sigma
  d\Delta\phi_2(n_t,n_{\tbar}))$ (related to the TP
  $\epsilon_2=\epsilon(Q,\tbar,n_t,n_{\tbar})$), with the $\Delta
  R_{\ell\ell}$ cut turned off. As was the case in table~\ref{table3},
  the sign of the parameter $a_1$ changes for
  $\kp=0,\kp=1$ and we restrict $\delta$ to be between
    $\pm \pi/2$.}
\label{table4}
\begin{center}
\begin{tabular}{|C{1cm}|C{1cm}||C{3cm}|C{3cm}|C{3cm}|}
%\begin{tabular}{|c|r||r|c||r|c||r|c|}
\hhline{|=====|}
%\hhline{|--------|}
$\kappa_t$&$\tilde{\kappa}_t$~~&$a_0$~~&$a_1$& $\delta$~~ \\ 
\hhline{|=====|} 
%\hhline{|--------|}
$1$ & $-1$~~~ & $0.1591 \pm 0.0006$ & $-0.0146 \pm 0.0008$ & $0.73 \pm 0.06$ \\[0.6mm]
\hline
$1$ & $0$ & $0.1594 \pm 0.0007$ & $-0.0190 \pm 0.0008$ & $0.005 \pm 0.06\,\,$ \\[0.6mm]
\hline
$1$ & $1$ & $0.1592 \pm 0.0006$ & $-0.0136 \pm 0.0008 $ & $\,-0.77 \pm 0.07\quad$  \\[0.6mm]
\hline
$0$ & $1$ & $0.1591 \pm 0.0006$ & ~~$\,0.0113 \pm 0.0008$ & $0.09 \pm 0.08$ \\
\hhline{|=====|}
%\hhline{|--------|}
\end{tabular}
\end{center} 
\end{table}
%%%%%%%%%%%%%%%%%%%%%%%%%%%%%%%%%%%%%%%%%%%%%%%%%%%%%%%%%%%%%%%%%%%%%%%%%%
\subsection{Mean value}
\label{sec3.3}
We turn now to consider the last type of observable that we will
construct from the TPs, the mean value. As was the case for the
observables considered in sections~\ref{sec3.1} and \ref{sec3.2},
the mean value is sensitive to $\kpt$.  Given a certain
TP, we define its mean value in the following manner,
\beq
\label{eq28}
\langle \epsilon \rangle = \frac{\int\epsilon\, [d\sigma (pp\to b\,\ell^+\nu_{\ell}\,\bbar\,\ell^-\bar{\nu}_{\ell}H)/ d\Phi]\,d\Phi}{\int [d\sigma (pp\to b\,\ell^+\nu_{\ell}\,\bbar\,\ell^-\bar{\nu}_{\ell}H)/ d\Phi]\,d\Phi},
\eeq
where $\Phi$ is the Lorentz-invariant phase space corresponding to the
final state
$b\,\ell^+\nu_{\ell}\,\bbar\,\ell^-\bar{\nu}_{\ell}H$. From
eq.~(\ref{eq20}) we see that only the terms linear in (both) $\kp$ and $\kpt$
will contribute to the mean value.  Thus, we expect this observable
to be sensitive not only to the magnitude of $\kp\kpt$,
but also to the relative sign of
the couplings.\par

The results obtained for the TPs $\epsilon_1=
\epsilon(t,\tbar,n_t,n_{\tbar})$, $\epsilon_2=
\epsilon(Q,\tbar,n_t,n_{\tbar})$ and $\epsilon_3
=\epsilon(Q,t,n_t,n_{\tbar})$ introduced in section~\ref{sec2} are
displayed in table~\ref{table5}.  For each TP we list the 
mean value divided by the
corresponding statistical uncertainty.
%A deviation  of the estimator of $\langle
%\epsilon \rangle$, $\bar{\epsilon}$.
We see that the three observables are capable of distinguishing the SM
case from both $\mathrm{CP}$-mixed cases. Furthermore,
the two $\mathrm{CP}$-mixed cases
are clearly disentangled, since the observables are
sensitive to the sign of $\kpt$. The observables $\langle \epsilon_2
\rangle$ and $\langle \epsilon_3 \rangle$ appear to be slightly more
sensitive than $\langle \epsilon_1 \rangle$.  Also, the
mean value for the combination $\epsilon_4$ introduced in
section~\ref{sec3.1} is slightly less sensitive than $\langle \epsilon_1\rangle$,
 $\langle \epsilon_2\rangle$ and $\langle \epsilon_3\rangle$, 
 with values $-4.32$, $1.11$ and $7.23$ for the cases
$(\kp=1,\kpt=-1,0,1)$, respectively. As with the asymmetry, the
purely $\mathrm{CP}$-even and purely $\mathrm{CP}$-odd cases cannot be
distinguished by the mean value, since it is linear in both $\kp$ and
$\kpt$ (see eqs.~(\ref{eq20}) and (\ref{eq28})). Comparing the results
in table~\ref{table5} with
the results presented in section~\ref{sec3.1}, we can conclude that
the sensitivity to the anomalous $tH$ coupling is smaller for the mean values
of the TPs under consideration
than for the corresponding asymmetries.
\renewcommand{\arraystretch}{1.4}
\begin{table}[H]
\caption{Mean values obtained for the TPs $\epsilon_{1,2,3}$ for the
  SM case and two $\mathrm{CP}$-mixed cases.
  The values are obtained using a sample of $10^5$
  simulated events.}
\label{table5}
\begin{center}
\begin{tabular}{|C{1cm}|C{1cm}||C{2cm}||C{2cm}||C{2cm}|}
%\hhline{|=====|}
\hhline{|=====|}
$\kappa_t$ & $\tilde{\kappa}_t$ & $\langle \epsilon_1 \rangle /\sigma_{\bar{\epsilon}_1}$ & $\langle \epsilon_2 \rangle /\sigma_{\bar{\epsilon}_2}$ & $\langle \epsilon_3 \rangle /\sigma_{\bar{\epsilon}_3}$ \\ 
\hhline{|=====|} 
%\hhline{|-----|}
%\vspace*{.5mm}
\renewcommand{\arraystretch}{1.0}
$1$~ & $-1$~~~ & $4.26$~ & $4.94$~ & $-5.81$~~~\\[0.6mm]
\hline
$1$ & $0$~ & $-0.91$~~~ & $-0.22$~~~ & $1.25$~\\[0.6mm]
\hline
$1$ & $1$~ & $-7.98$~~~ & $-8.83$~~~& $8.75$~\\[0.6mm]
\hhline{|=====|}
%\hhline{|-----|}
\end{tabular}
\end{center} 
\end{table}
%%%%%%%%%%%%%%%%%%%%%%%%%%%%%%%%%%%%%%%%%%%%%%%%% 
\section{{\boldmath $\mathrm{CP}$}-odd observables not depending on {\boldmath $t$} and {\boldmath $\tbar$} spin vectors}
\label{sec4}
So far we have considered TPs involving the momenta $t,\tbar$
and $Q$ and the spin vectors $n_t$ and $n_{\tbar}$ [defined in
eqs.~(\ref{eq3})-(\ref{eq4})].  Furthermore, we have described the general
form of the differential cross section in terms of these vectors in
eq.~(\ref{eq20}). In this section we consider other
possibilities for the choice of the vectors from which the
$\mathrm{CP}$-odd observables can be constructed. From the definitions
in eqs.~(\ref{eq3}) and (\ref{eq4}), we see that the TPs
$\epsilon_{1,2,3}$ can be written as follows,
\beq
\label{eq29}
\TPa = \frac{m^2_t}{(t\cdot \ell^+)(\tbar\cdot \ell^-)}\,\epsilon(t,\tbar,\ell^-\!,\ell^+),
\eeq
\vspace*{1mm}
\beq
\label{eq30}
\TPb = \frac{m^2_t}{(t\cdot \ell^+)(\tbar\cdot \ell^-)}\left(\epsilon(t,\tbar,\ell^-\!,\ell^+)+\epsilon(H,\tbar,\ell^-\!,\ell^+)+\frac{(t\cdot\ell^+)}{m^2_t}\epsilon(H,\tbar,t,\ell^-\!)\right) \,,
\eeq
\vspace*{1mm}
\beq
\label{eq31}
\TPc = \frac{m^2_t}{(t\cdot \ell^+)(\tbar\cdot \ell^-)}\left(-\epsilon(t,\tbar,\ell^-\!,\ell^+)+\epsilon(H,t,\ell^-\!,\ell^+)+\frac{(\tbar\cdot\ell^-)}{m^2_t}\epsilon(H,\tbar,t,\ell^+\!)\right).
\vspace*{2mm}
\eeq
The above equations express the TPs studied in the last sections as
a combination of TPs involving the momenta $t,\tbar,H,\ell^+$ and
$\ell^-$, with coefficients that are functions of phase space
variables. These five momenta give rise to five TPs whose sensitivity
can also be tested by means of the observables introduced in
sections~\ref{sec3.1}-\ref{sec3.3}.  We have found
that TPs that do not include both the lepton and anti-lepton momenta yield
negligible sensitivity to the value of $\kpt$.  For this reason,
we concentrate here on the
results obtained for the remaining TPs,\footnote{These TPs should not be confused with those introduced in eq.~(\ref{eq20}).}
\beq
\label{eqa5}
\epsilon_5\equiv\epsilon(t,\tbar,\ell^-\!,\ell^+) \,,
\vspace*{1mm}
\eeq
\beq
\label{eqa6}
\epsilon_6\equiv\epsilon(H,t,\ell^-\!,\ell^+) \,,
\vspace*{1mm}
\eeq
\beq
\label{eqa7}
\epsilon_7\equiv\epsilon(H,\tbar,\ell^-\!,\ell^+) \,.
\vspace*{1mm}
\eeq
\par
tables~\ref{table7} and
\ref{table6} summarize the results for the TPs $\epsilon_{5,6,7}$.
We see that $\epsilon_5$ gives rise to asymmetries and mean
values that are clearly larger than those obtained for $\epsilon_6$
and $\epsilon_7$. This is in contrast to the TPs $\epsilon_{1,2,3}$,
for which the asymmetries and mean values are comparable among the
TPs
(see tables
\ref{table1} and \ref{table5}). We also note that the asymmetry for
$\epsilon_5$ is exactly the same as for $\epsilon_1$, as is expected
from eq.~(\ref{eq29}), since the proportionality factor relating them
is positive definite. Regarding the mean values, we see by comparing
tables~\ref{table5} and \ref{table6} that the TPs $\epsilon_{1,2,3}$
appear to have a higher sensitivity to the pseudoscalar coupling than do
$\epsilon_{5,6,7}$. 
\renewcommand{\arraystretch}{1.4}
\begin{table}[H]
\caption{Asymmetries for the TPs $\epsilon_{5,6,7}$ for the SM case
  and the two $\mathrm{CP}$-mixed cases.
  The values correspond to $10^5$ simulated events.}
\label{table7}
\begin{center}
\begin{tabular}{|C{0.8cm}|C{0.8cm}||C{1.9cm}|C{1.9cm}||C{1.9cm}|C{1.9cm}||C{1.9cm}|C{1.9cm}|}
%\begin{tabular}{|c|r||r|c||r|c||r|c|}
\hhline{|========|}
%\hhline{|--------|}
$\kappa_t$&$\tilde{\kappa}_t$~~&$\mathcal{A}(\epsilon_5)$~~&$\mathcal{A}(\epsilon_5)/\sigma_{\mathcal{A}}$& $\mathcal{A}(\epsilon_6)$~~&$\mathcal{A}(\epsilon_6)/\sigma_{\mathcal{A}}$&$\mathcal{A}(\epsilon_7)$~~&$\mathcal{A}(\epsilon_7)/\sigma_{\mathcal{A}}$  \\ 
\hhline{|========|} 
%\hhline{|--------|}
$1$ & $-1$~~~ & $0.0315$~ & $10.0$~ & $-0.0134$~~~ & $-4.2$~~~ & $0.0111$~ & $3.5$~\\[0.6mm]
\hline
$1$ & $0$ & $-0.0021$~~~ & $-0.7$~~~ & $-0.0011$~~~ & $-0.3$~~~ & $0.0009$~& $0.3$~\\[0.6mm]
\hline
$1$ & $1$ & $-0.0379$~~~ & $-12.0$~~~ & $0.0143$~ & $4.5$~ & $-0.0137$~~~ & $\,-4.3$~~~  \\[0.6mm]
\hhline{|========|}
%\hhline{|--------|}
\end{tabular}
\end{center} 
\end{table}
\renewcommand{\arraystretch}{1.4}
\begin{table}[H]
\caption{Mean values obtained for $\epsilon_{5,6,7}$ for the
  SM case and the two $\mathrm{CP}$-mixed cases.
  The values correspond to $10^5$ simulated
  events.}
\label{table6}
\begin{center}
\begin{tabular}{|C{1cm}|C{1cm}||C{2cm}||C{2cm}||C{2cm}|}
\hhline{|=====|}
%\hhline{|-----|}
$\kappa_t$ & $\tilde{\kappa}_t$ & $\langle \epsilon_5 \rangle /\sigma_{\bar{\epsilon}_5}$ & $\langle \epsilon_6 \rangle /\sigma_{\bar{\epsilon}_6}$ & $\langle \epsilon_7 \rangle /\sigma_{\bar{\epsilon}_7}$ \\ 
\hhline{|=====|} 
%\hhline{|-----|}
%\vspace*{.5mm}
\renewcommand{\arraystretch}{1.0}
$1$~ & $-1$~~~ & $3.98$~ & $-1.96$~~~ & $1.69$~ \\[0.6mm]
\hline
$1$ & $0$~ & $-0.43$~~~ & $1.25$~ & $0.74$~ \\[0.6mm]
\hline
$1$ & $1$~ & $-6.76$~~~ & $3.46$~ & $-3.29$~~~~\\[0.6mm]
\hhline{|=====|}
%\hhline{|-----|}
\end{tabular}
\end{center} 
\end{table}
 \par
It is important to mention that in the
$t\tbar$ rest frame the sign of the TP $\epsilon_5$ is defined through
the angle $\Delta\phi(\ell^-,\ell^+)$ (see the discussion following
  eq.~(\ref{eq25})),
which is the angular difference between the
projections of the leptons' momenta onto the plane perpendicular to
$\vec{\tbar}$.  As in section~\ref{sec3.2}, we can construct an associated angular
distribution (see eq.~(\ref{eq26})) that
will be sensitive to the sign of the pseudoscalar coupling. The
angular variable $\Delta\phi(\ell^-,\ell^+)$ 
is the same as that proposed in ref.~\cite{Ellis} as a
useful $\mathrm{CP}$-odd observable. Moreover, it is shown in
ref.~\cite{Ellis} that the corresponding
angular distribution follows the functional
form given in eq.~(\ref{eq27}). The associated shifts ($\delta$)
obtained for different values of $\kpt$ are expected to be of the same
order as those exhibited by the $\Delta\phi_1(n_t,n_{\tbar})$
distribution since the $\Delta\phi(\ell^-,\ell^+)$ distribution 
is constrained by the asymmetry $\mathcal{A}(\epsilon_5)$
[via eq.~(\ref{eq26})],
which in turn is equal to
$\mathcal{A}(\epsilon_1)$. Also, we note that $\mathcal{A}(\epsilon_5)$ is slightly less sensitive
than $\mathcal{A}(\epsilon_2)$,
as can be seen from table~\ref{table1}.
\par

In addition to the $\epsilon_5$ angular distribution
(defined above), one can also define angular distributions
corresponding to $\epsilon_6$ and $\epsilon_7$.  As was the case
for the $\epsilon_5$ distribution,
the corresponding angles will be defined in terms
of the momenta of the leptons instead of in terms of the spin vectors
(as was done in section~\ref{sec3.2}). The
angular distributions based on $\epsilon_5$-$\epsilon_7$
have the same overall behaviour as those derived from
$\epsilon_1$-$\epsilon_3$.  Using eq.~(\ref{eq27})
to fit the distributions and comparing to the results obtained for
$\epsilon_1$-$\epsilon_3$, we find that the phase shifts ($\delta$)
are comparable for the $\epsilon_5$ angular distribution, but are
smaller for the
$\epsilon_6$ and $\epsilon_7$ distributions.  
\par

In analogy with the
combination of TPs considered in section~\ref{sec3}, we have found a
combination of the TPs $\epsilon_{5,6,7}$ for which the asymmetry is
enhanced compared to those for $\epsilon_5$-$\epsilon_7$,
\beq
\label{eq32}
\epsilon_8 = 2\epsilon_5 -\epsilon_6 +\epsilon_7 = \epsilon(t+\tbar+H,t-\tbar,\ell^+,\ell^-).
%2\epsilon(t,\tbar,\ell^-,\ell^+)+\epsilon(H,\tbar,\ell^-,\ell^+)-\epsilon(H,t,\ell^-,\ell^+)
\eeq
We see from eq.~(\ref{eq32}) that in the $t\tbar H$ rest frame
$\epsilon_8=M_{t\tbar H}(\vec{t}-\vec{\tbar})\cdot (\vec{\ell}^+\!
\times \vec{\ell}^-)$, where $M_{t\tbar H}$ is the invariant mass of
the $t\tbar H$ system. Hence, in the $t\tbar H$ rest frame the sign of 
$\epsilon_8$ is determined by the quantity
$(\vec{t}-\vec{\tbar})\cdot (\vec{\ell}^+\! \times
\vec{\ell}^-)$. Comparing eqs.~(\ref{eq24}) and (\ref{eq32}),
and noting that $Q=(t+\tbar+H)/2$, we see that the only
relevant difference between $\epsilon_4$ and $\epsilon_8$ is that in
the latter the spin vectors $n_t$ and $n_{\tbar}$ have been replaced
by the momenta of the leptons $\ell^+$ and $\ell^-$, respectively. The
values obtained for $\mathcal{A}(\epsilon_8)$ are shown in table
\ref{table8}. Compared to the TPs $\epsilon_1$-$\epsilon_3$ and
$\epsilon_5$-$\epsilon_7$ (see tables \ref{table1} and \ref{table7}),
the asymmetry for $\epsilon_8$ has a comparable or slightly higher
sensitivity for resolving the $\mathrm{CP}$-mixed cases.
Comparing with $\mathcal{A}(\epsilon_4)$, however,
we see that using the momenta of the leptons (in $\epsilon_8$)
instead of the spin vectors produces a decrease in the sensitivity of
the asymmetry (see tables \ref{table2} and
\ref{table8}).

The mean values of $\epsilon_8$ for the scenarios under
consideration are comparable with the values listed in table
\ref{table6} for $\epsilon_5$. 
We have also studied the associated angular distributions. Specifically, 
we have analyzed the $\Delta\phi(\ell^+,\ell^-)$ distribution in the $t\tbar H$ rest frame,
taking $H$ to define the $z$-axis. The distributions obtained for different values of
$\kp$ and $\kpt$ are not well described by eq.~(\ref{eq27}) (the situation is similar
to that encountered for the angular distribution
associated with $\epsilon_4$ -- see the discussion
at the end of section~\ref{sec3.2}.).  For different values of the parameters,
the distributions become slightly distorted giving rise to a non-zero asymmetry 
(see eq. (29)).
\begin{table}[H]
\caption{Asymmetry for the TP $\epsilon_{8}$ for the SM case and the
  two $\mathrm{CP}$-mixed scenarios. The
  values are obtained with $10^5$ simulated events.}
\label{table8}
\begin{center}
\begin{tabular}{|C{1cm}|C{1cm}||C{2cm}|C{2cm}|}
%\begin{tabular}{|c|r||r|c||r|c||r|c|}
\hhline{|====|}
%\hhline{|--------|}
$\kappa_t$&$\tilde{\kappa}_t$~~&$\mathcal{A}(\epsilon_8)$~~&$\mathcal{A}(\epsilon_8)/\sigma_{\mathcal{A}}$ \\ 
\hhline{|====|} 
%\hhline{|--------|}
$1$ & $-1$~~~ & $0.0331$~ & $10.5$~ \\[0.6mm]
\hline
$1$ & $0$ & $0.0023$~ & $0.7$~ \\[0.6mm]
\hline
$1$ & $1$ & $-0.0403$~~~ & $\,-12.7$~~~~ \\[0.6mm]
\hhline{|====|}
%\hhline{|--------|}
\end{tabular}
\end{center} 
\end{table}
\section{{\boldmath{$\mathrm{CP}$}}-odd observables not depending on {\boldmath $t$} and {\boldmath $\tbar$} momenta}
\label{sec5}
The observables discussed in the preceding sections all involve the
momenta of the top and/or anti-top quarks and thus require the full
reconstruction of the kinematics of the individual $t$ and $\tbar$
systems in order to be measured. Although challenging due to the
presence of the two neutrinos in the final state, this can in principle be
done by applying a kinematic reconstruction algorithm (we will come back to this point in the next section). Another
possibility is to define observables that do not depend on the $t$
and $\tbar$ momenta but instead make use of the momenta of the
$b$ and $\bbar$ quarks to which the $t$ and $\tbar$ decay. In
order to construct such observables we will take as our starting
point the most
sensitive observables studied in sections~\ref{sec3} and \ref{sec4},
namely those associated with the TPs $\epsilon_4$ and $\epsilon_8$,
respectively.

Let us first consider the TP combination $\epsilon_8$, which is
defined in eq.~(\ref{eq32}).  Replacing the momenta of the
$t$ and $\tbar$ quarks by the momenta of the $b$ and $\bbar$ quarks,
respectively, we have a new TP,
\beq
\label{eq33}
\epsilon_9 = \epsilon(b+\bbar+H,b-\bbar,\ell^+,\ell^-).
\eeq
Note that the sign of $\epsilon_9$ is determined by the sign of
the quantity $(\vec{b}-\vec{\bbar})\cdot(\vec{\ell}^+\!\times
\vec{\ell}^-)$ in the $b\bbar H$ rest frame. This combination
of three vectors (determined in the lab frame instead
of the $b\bbar H$ rest frame) is
used in ref.~\cite{Guadagnoli} to define a
$\mathrm{CP}$-odd observable that only depends on lab frame
variables. The values of the asymmetry for $\epsilon_9$ are listed in
table~\ref{table9}. Comparing tables~\ref{table8} and \ref{table9}
we see that the use of the $b$ and $\bbar$ momenta
instead of the $t$ and $\tbar$ momenta leads to a
decrease in the sensitivity of the asymmetry by $\sim 5\sigma$ for
$\kp=1,\kpt=\pm 1$. Nevertheless, the observable can still
discriminate not only between the two $\mathrm{CP}$-mixed scenarios
but also between these and the SM case.
%\newpage

%
\begin{table}[H]
\caption{Asymmetry for the TP $\epsilon_{9}$ for the SM case and the
  two $\mathrm{CP}$-mixed cases. The
  values are obtained with $10^5$ simulated events.
  }
\label{table9}
\begin{center}
\begin{tabular}{|C{1cm}|C{1cm}||C{2cm}|C{2cm}|}
%\begin{tabular}{|c|r||r|c||r|c||r|c|}
\hhline{|====|}
%\hhline{|--------|}
$\kappa_t$&$\tilde{\kappa}_t$~~&$\mathcal{A}(\epsilon_9)$~~&$\mathcal{A}(\epsilon_9)/\sigma_{\mathcal{A}}$ \\ 
\hhline{|====|} 
%\hhline{|--------|}
$1$ & $-1$~~~ & $0.0171$~ & $5.4$~ \\[0.6mm]
\hline
$1$ & $0$ & $0.0010$~ & $0.3$~ \\[0.6mm]
\hline
$1$ & $1$ & $-0.0247$~~~ & $\,-7.8$~~~~ \\[0.6mm]
\hhline{|====|}
%\hhline{|--------|}
\end{tabular}
\end{center} 
\end{table}

We proceed in a similar manner with the TP $\epsilon_4$. Starting from
eq.~(\ref{eq24}) and using the definitions of the spin vectors
in eqs.~(\ref{eq3}) and (\ref{eq4}), we have
\beq
\label{eq34}
\epsilon_4 = \frac{m^2_t}{(t\cdot\ell^+)\cdot(\tbar\cdot\ell^-)}\,\epsilon(Q,t-\tbar,\ell^-,\ell^+)+\frac{1}{(t\cdot\ell^+)}\,\epsilon(Q,t,\ell^+,\tbar)-\frac{1}{(\tbar\cdot \ell^-)}\,\epsilon(Q,\tbar,t,\ell^-).
\eeq
Since the asymmetry is not changed by the presence
of an overall positive definite multiplicative
factor, let us concentrate instead on the following combination
of TPs,
\beq
\label{eq35}
\epsilon(Q,t-\tbar,\ell^-,\ell^+)+\frac{(\tbar\cdot \ell^-)}{m^2_t}\epsilon(Q,t,\ell^+,\tbar)-\frac{(t\cdot\ell^+)}{m^2_t}\epsilon(Q,\tbar,t,\ell^-).
\eeq
Instead of replacing $t$ and $\tbar$ directly by $b$ and $\bbar$,
we use the visible contributions, namely $b+\ell^+$ and $\bbar +\ell^-$,
respectively. This results in the following definition
\beq
\label{eq36}
\epsilon_{10}=\epsilon(\tilde{Q},c_{b\bbar}\,,\ell^-,\ell^+)-w_1\,\epsilon(\tilde{Q},b,\bbar,\ell^+)+w_2\,\epsilon(\tilde{Q},b,\bbar,\ell^-),
\eeq
where $\tilde{Q}\equiv (b+\ell^+\!+\bbar +\ell^-)/2$ stands for the
visible part of $Q$, $c_{b\bbar}=(1-w_1)\,b-(1-w_2)\,\bbar$, and the
weights $w_{1,2}$ are given by $(\bbar\cdot \ell^-)/m^2_t $ and
$(b\cdot \ell^+)/m^2_t$, respectively. Also, the contribution
$m^2_{\ell}/m^2_t$ has been neglected both in $w_1$ and in $w_2$. Note
that if we set $w_1=w_2=0$, the combination $\epsilon_{10}$ reduces to
$\epsilon_9 /2$ and $\mathcal{A}(\epsilon_{10})$ becomes equal to
$\mathcal{A}(\epsilon_9)$. The results obtained for the asymmetry of
$\epsilon_{10}$ are given in table \ref{table10}. By comparing tables
\ref{table2} and \ref{table10} we see again that the sensitivity of
the asymmetry decreases when $t$ and $\tbar$ are not included in the
TP. Nevertheless, the combination $\epsilon_{10}$ remains a useful
observable for discriminating the $\mathrm{CP}$ nature of the Higgs
boson, with the corresponding asymmetry having a sensitivity that is
higher than that of $\epsilon_9$.

Comparing
tables~\ref{table9} and \ref{table10}, we see that the effective
separation between the $\mathrm{CP}$-mixed scenarios is enhanced by
about $3\sigma$ for $\mathcal{A}(\epsilon_{10})$ compared
  to $\mathcal{A}(\epsilon_{9})$.
    This improvement in the asymmetry may be due to two facts. In the first
place, as was pointed out in section~\ref{sec4} when comparing the TPs
$\epsilon_4$ and $\epsilon_8$, the asymmetry appears to be higher when
the spin vectors are used instead of the lepton momenta.  We see
from eqs.~(\ref{eq33}) and (\ref{eq36}) that $\epsilon_{10}$, being
obtained from $\epsilon_4$, contains the information on the spin
vectors; by way of contrast, $\epsilon_9$ depends directly on the
lepton momenta because it is derived from $\epsilon_8$. In the second
place, in order to obtain $\epsilon_{10}$, we have replaced
the top and antitop momenta by their 
visible parts, while in the case of $\epsilon_9$ the bottom and
antibottom momenta have been used.\par
For comparison purposes, we have also used our simulated events
to test the lab frame
\begin{table}[H]
\caption{Asymmetry for the TP $\epsilon_{10}$ for the SM case and the
  two $\mathrm{CP}$-mixed cases. The
  values are obtained by using $10^5$ simulated events.}
\label{table10}
\begin{center}
\begin{tabular}{|C{1cm}|C{1cm}||C{2cm}|C{2cm}|}
%\begin{tabular}{|c|r||r|c||r|c||r|c|}
\hhline{|====|}
%\hhline{|--------|}
$\kappa_t$&$\tilde{\kappa}_t$~~&$\mathcal{A}(\epsilon_{10})$~~&$\mathcal{A}(\epsilon_{10})/\sigma_{\mathcal{A}}$ \\ 
\hhline{|====|} 
%\hhline{|--------|}
$1$ & $-1$~~~ & $-0.0213$~~~ & $-6.7$~~~ \\[0.6mm]
\hline
$1$ & $0$ & $0.0031$~ & $1.0$~ \\[0.6mm]
\hline
$1$ & $1$ & $0.0300$~ & $9.5$~ \\[0.6mm]
\hhline{|====|}
%\hhline{|--------|}
\end{tabular}
\end{center} 
\end{table}
\noindent
observable given
in ref.~\cite{Guadagnoli}. We have found that this observable
appears to
be slightly less sensitive than $\mathcal{A}(\epsilon_{10})$,
giving rise to an effective separation between the $\mathrm{CP}$-mixed scenarios
that is smaller by about $1.4\sigma$.

\section{Experimental Feasibility}
\label{sec6}

In our numerical analyses so far we have used relatively large
samples of events ($10^5$ events per sample)
in order to clearly distinguish
which observables would be most promising.  
The number of events expected at the High Luminosity Large
Hadron Collider (HL-LHC), however, is
smaller than the number of events that we have used in our
simulations.  In this section we reexamine
the more promising observables, using sample sizes that are
more attainable in the near future.  

Let us first make some estimates regarding the number of signal events
expected at the HL-LHC.
In section~\ref{sec3} we introduced several mild selection cuts.
Implementing these cuts, and assuming that the final state leptons
could be either electrons or muons, the SM cross section for $\ppprocess$ at
$14\,\mathrm{TeV}$ is $\sim 15.3\,\mathrm{fb}$; thus, the number of
events expected within the context of the HL-LHC is $\sim
15.3\,\mathrm{fb} \times 3000\,\mathrm{fb}^{-1} = 4.59\times
10^4$. This number is expected to be larger
if $\kpt\neq 0$ (assuming $\kp=1$),
since the corresponding cross section is
larger than the SM cross section in this case.
Taking into account NLO corrections (to the
production process) via a $K$ factor of approximately
$1.2$~\cite{Dawson,Beenakker,Dittmaier}, we find that
the expected number of
events increases to $\sim 5.5 \times 10^4$.
On the other hand, additional cuts, as
well as a reduction in efficiency related to momentum reconstruction,
will lead to a decrease in this number. \par

Given the discussion in the previous paragraph, we
have generated sets of $5\times 10^4, 1\times 10^4$ and $5\times 10^3$
events and have recalculated the most sensitive observable, 
$\mathcal{A}(\epsilon_4)$, for each case. The results are displayed in
table \ref{table11}, where it can be seen that for $5\times 10^4$ events
(which is close to our rough estimate above for the
total number of signal events for the
HL-LHC), 
the observable is still very sensitive to $\kpt$.
In this case, the $\mathrm{CP}$-mixed scenarios are
effectively separated by $19\sigma$. As expected, the
sensitivity worsens as the number of events is reduced, but even with
$5\times 10^3$ events the effective separation between the $\mathrm{CP}$-mixed
scenarios under consideration is $6.5\sigma$. \par

In section~\ref{sec5} we defined the TP combination $\epsilon_{10}$,
which does not depend directly on the top or antitop momenta.
Although the top and antitop momenta would not need to be
reconstructed to measure $\mathcal{A}(\epsilon_{10})$,
it is still useful to examine this observable for more
conservative numbers of events. table~\ref{table12} shows the
results obtained for 
$5\times 10^4$ and $1\times 10^4$ events. We see in this case
that even with $1\times 10^4$ events the observable is
able to distinguish the $\mathrm{CP}$-mixed cases by
$5.6\sigma$. 
\newcolumntype{D}[1]{>{\centering\arraybackslash}p{#1}}
\begin{table}[H]
\caption{Asymmetry for the TP $\epsilon_4$ obtained using $5\times
  10^4, 1 \times 10^4$ and $5\times 10^3$ events for the SM
  case and the two $\mathrm{CP}$-mixed cases. }
\label{table11}
\begin{center}
\begin{tabular}{|D{0.8cm}|D{0.8cm}||D{1.9cm}|D{1.9cm}||D{1.9cm}|D{1.9cm}||D{1.9cm}|D{1.9cm}|}
%\begin{tabular}{|c|r||r|c||r|c||r|c|}
\hhline{|========|}
%\hhline{|--------|}
\multirow{2}{*}{$\kappa_t$} & \multirow{2}{*}{$\tilde{\kappa}_t$} & \multicolumn{2}{c||}{$N_{\mathrm{ev}}=5\times 10^4$} & \multicolumn{2}{c||}{$N_{\mathrm{ev}}=1\times 10^4$} & \multicolumn{2}{c|}{$N_{\mathrm{ev}}=5\times 10^3$} \\ \cline{3-8}
& & $\mathcal{A}(\epsilon_4)$~~&$\mathcal{A}(\epsilon_4)/\sigma_{\mathcal{A}}$ &  $\mathcal{A}(\epsilon_4)$~~&$\mathcal{A}(\epsilon_4)/\sigma_{\mathcal{A}}$ &  $\mathcal{A}(\epsilon_4)$~~&$\mathcal{A}(\epsilon_4)/\sigma_{\mathcal{A}}$ \\
\hhline{|========|} 
%\hhline{|--------|}
$1$ & $-1$~~~ & $-0.0405$~~~ & $-9.1$~~~ & $-0.0426$~~~ & $-4.3$~~~ & $-0.0496$~~~ & $-3.5$~~~ \\[0.6mm]
\hline
$1$ & $0$ & $0.0004$~ & $0.1$~ & $-0.0084$~~~ & $-0.8$~~~ & $-0.0004$~~~ & $-0.03$~~~ \\[0.6mm]
\hline
$1$ & $1$ & $0.0443$~ & $9.9$~ & $0.0434$~ & $4.2$~ & $\,0.0420$~ & $3.0\,\,$ \\
\hhline{|========|}
%\hhline{|--------|}
\end{tabular}
\end{center} 
\end{table}
\begin{table}[H]
\caption{Asymmetry for the TP $\epsilon_{10}$ in the SM case and the
  two $\mathrm{CP}$-mixed cases for
  $5\times 10^4$ and $1\times 10^4$ events.}
\label{table12}
\begin{center}
\begin{tabular}{|C{1cm}|C{1cm}||C{2cm}|C{2cm}||C{2cm}|C{2cm}|}
%\begin{tabular}{|c|r||r|c||r|c||r|c|}
\hhline{|======|}
%\hhline{|--------|}
\multirow{2}{*}{$\kappa_t$} & \multirow{2}{*}{$\tilde{\kappa}_t$} & \multicolumn{2}{c||}{$N_{\mathrm{ev}}=5\times 10^4$} & \multicolumn{2}{c|}{$N_{\mathrm{ev}}=1\times 10^4$} \\ \cline{3-6}
& & $\mathcal{A}(\epsilon_{10})$~~&$\mathcal{A}(\epsilon_{10})/\sigma_{\mathcal{A}}$ &  $\mathcal{A}(\epsilon_{10})$~~&$\mathcal{A}(\epsilon_{10})/\sigma_{\mathcal{A}}$ \\
\hhline{|======|} 
%\hhline{|--------|}
$1$ & $-1$~~~ & $-0.0270$~~~ & $-6.0$~~~ & $-0.0184$~~~ & $-1.8$~~~  \\[0.6mm]
\hline
$1$ & $0$ & $0.0022$~ & $0.5$~ & $-0.0086$~~~ & $-0.9$~~~  \\[0.6mm]
\hline
$1$ & $1$ & $0.0313$~ & $7.0$~ & $0.0380$~ & $3.8\,$ \\
\hhline{|======|}
%\hhline{|--------|}
\end{tabular}
\end{center} 
\end{table}
%
 % \noindent 
We note that in order to be fully conclusive about the required luminosity, it is important to include the effects of hadronization,
detector resolution, reconstruction efficiencies and so
forth. In fact, the measurement of some of the proposed observables
necessitates the reconstruction of the $t$ and $\tbar$
momenta. Such is the case, for example, for the most sensitive observable,
the asymmetry $\mathcal{A}(\epsilon_4)$.

The determination of
the kinematic quantities associated with the top quark and antiquark is
challenging, not only due to the presence of the two neutrinos in the
final state (which escape the detector undetected), but also because
the (visible) quarks and charged leptons in the final state need to be
correctly associated with the corresponding parent particle (i.e., the
top or antitop quark). Even in the case in which two leptons and two
jets are reconstructed, there are still two possibilities for associating
the $b$ jets with the appropriate parent particles. Regarding
the momenta of the neutrinos, the six unknowns
(corresponding to the three-momenta of the two neutrinos) can be
determined by using the six kinematic equations following from the 
conservation of the transverse momentum
and from the $W^\pm$ and $t$ and $\bar{t}$ invariant mass constraints.
As is shown in ref.~\cite{Sonnenschein},
the resulting set of equations can be reduced to one univariate polynomial of
degree four, leading to the possibility of obtaining up to four
solutions. In addition to these various challenges, the impact of the
finite detector resolution on finding the solution of the kinematic
equations has to be taken into account. There are various methods
of kinematic reconstruction that deal with these problems and allow for
the reconstruction of the kinematical properties of the top-quark pair
from the four-momenta of the final-state particles. The following
describes two kinematic reconstruction methods used recently by the
ATLAS and CMS collaborations.
\begin{itemize}
\item The first method is known as the neutrino weighting technique and is
  based on ref.~\cite{Abbott}. In this approach, the kinematic equations
  are used with the reconstructed jets, leptons and $\vmet$ as inputs
  and the masses of the $W$ bosons, the $t$ and the $\tbar$ are fixed. The
  pseudorapidities corresponding to the two neutrinos are sampled by
  using a simulated neutrino energy spectrum and, in order to include
  detector resolution effects, the reconstructed jets are
  smeared. Each solution obtained by scanning over the two
  pseudorapidities for each smearing step are weighted according to
  the agreement between the calculated and measured $\vmet$. For each
  event, the measurement of a given observable is obtained as the
  respective weighted mean value.  Within the context of $t\tbar$
  production this procedure has been used, for instance, to obtain
  spin correlation~\cite{atlasconf} and charge
  asymmetry~\cite{atlascharge} measurements in the dileptonic decay
  channel. In the former case, the reconstruction efficiency is approximately
  $95\%$ for simulated $t\tbar$ events, while in the latter case this
  efficiency is estimated to be $80\%$ for the experimental data set.
\item The second method also uses the kinematic equations with the
  reconstructed objects as inputs, but in contrast to the previous
  method, only the top quark mass is fixed (to the value
  $m_t=172.5\,\mathrm{GeV}$); the $W$ mass is smeared according
  to the true $W$ mass distribution.
  The energies and the
  directions of the reconstructed jets and leptons are smeared $100$
  times and events with two $b$-tagged jets are preferred compared to those
  with one $b$-tagged jet. For each lepton-jet pair a weight is
  assigned based on the expected true lepton-$b$-jet invariant mass
  spectrum, and the pair with the highest sum (over the smearings) of
  weights is chosen. For each of the $100$ smearings of this
  lepton-jet pair, the ambiguity in the solution of the kinematics
  equations is resolved 
  by taking the solution giving the smallest
  invariant mass of the $t\tbar$ system. Finally, the kinematic
  quantities associated with the top quark and antiquark are obtained as
  a weighted average according to the true $m_{b\ell}$
  distribution. This technique has been used in ref.~\cite{CMS2} to
  measure the differential cross-section for $t\tbar$ production in
  the dileptonic decay channel. The reconstruction efficiency reported
  is $\sim 94\%$, which is a $\sim 6\%$ improvement with respect to
  the method used in an earlier study on the same process~\cite{CMS1}.
\end{itemize}

In the case of $t\tbar H (H\rightarrow b\bbar)$ production at the LHC,
events reconstructed using the types of algorithms described above have
been used in the analysis of angular distributions that are useful for
discriminating the signal from the backgrounds~\cite{dosSantos}, with
the reconstruction efficiency being about $80\%$. The above kinematic
reconstruction algorithms proceed by using the reconstructed objects
as inputs. If the top quarks are produced with $p_T\sim
1\,\mathrm{TeV}$, the reconstruction of their decay products can be
complicated since they will be highly collimated. The application of
standard event reconstructions to the semileptonic decay of boosted
tops could lead to the merging of the corresponding $b$-jet and the
hard lepton. Moreover, the use of standard isolation requirements
leads to a low efficiency, which in turn depends on the top
polarization. A possibility for dealing with this problem is developed
in ref.~\cite{Tweedie}, where a set of baseline cuts that incorporate
a powerful isolation variable is used to recover the signal in the
muon channel. In particular, the use of this isolation variable allows
one to reject QCD jets with embedded leptons, and QCD jets in general, at
the level of $10^3$ and $10^4\sim 10^5$, respectively, while $80\sim
90\%$ of the tops are retained. Within the context of $pp$ collisions
at $\sqrt{s}=13\,\mathrm{TeV}$, the isolation criteria developed in
ref.~\cite{Tweedie} have been applied, for example, to experimental
searches for new heavy particles decaying into a pair of boosted tops
\cite{NoteAtlas1}, $t\tbar$ resonances decaying into semileptonic
boosted final states \cite{NoteCMS2}, $t\tbar H$ production in the
multilepton decay channel \cite{NoteCMS1} and four-top production in
the lepton+jets decay channel \cite{NoteAtlas2}, to name a
few analyzes.\footnote{Although the reconstruction technique of
  ref.~\cite{Tweedie} only considers the case of muons, it has also
  been applied to the case of electrons in
  refs.~\cite{NoteAtlas1,NoteAtlas2,NoteCMS1,NoteCMS2}.} \par
Finally, it is important to mention that a realistic analysis of the
sensitivity of the observables discussed in this paper also requires a
study of the impact of the backgrounds.  If we consider the dominant
decay mode of the Higgs boson, $H\rightarrow b\bbar$, in order to
maximize the cross section of the process, the signature is given by
$4$ $b$-jets, two leptons and missing energy.  The main background
arises from the production of $t\tbar$ in association with additional
jets, with the dominant source being the production of $t\tbar +
b\bbar$.  In ref.~\cite{chinos} it is shown that the application of a
small set of cuts results in a large improvement in the signal to
background ratio. On the experimental side, a rigorous treatment of
the signal and backgrounds for $t\tbar H$ production with $H\to
b\bbar$ is performed in ref.~\cite{atlasger}, using $20.3\,
\mathrm{fb}^{-1}$ of data at $\sqrt{s}=8\,\mathrm{TeV}$.
\par
In order to further study the most promising observables proposed in
this paper, it would be interesting
to perform a complete simulation, including the
hadronization and detector effects for the signal as well as for the
corresponding backgrounds, and then to apply the kinematic reconstruction
methods discussed above. However, this sort of analysis is beyond the
reach of the present study and is left as future work. Nevertheless,
the initial analysis performed in this paper paints an optimistic picture, since it indicates that the most sensitive observables proposed here can be
probed with a luminosity of order $300$-$600\,\mathrm{fb}^{-1}$,
which is attainable in the short term at the LHC.
%%%%%%%%%%%% with Cuts
%51.46/17 -0.02 +- 0.08
%45.48/17  -1.09 +- 0.07
%65.60/17   0.96 +- 0.08
%28.84/17   0.02 +- 0.07
%%%%%%%
%9.13/17     0.01 +- 0.06
%9.05/17     -0.98 +- 0.07
%19.7/17     0.80 +- 0.07
%16.59/17    0.06 +- 0.07
%
%%%%%%%%%%%% without Cuts
%16.36/17    0.0022 +-0.06
%15.25/17    -0.82 +- 0.07
%12.80/17    0.81 +- 0.07
%19.38/17    0.11 +- 0.08
%%%%%%
%7.45/17    0.005 +- 0.06
%18.17/17   -0.77 +- 0.07
%9.59/17    0.73 +- 0.06
%12.39/17   0.09 +- 0.08
%%%%%%%%%%%%%%%%%%%%%%%%%%%%%%%%%%%%%%%%%%%%%%%%%
\section{Conclusions}
\label{sec7}
In this paper we have presented a collection of $\mathrm{CP}$-odd
observables based on triple product correlations
in $\ppprocess$ that are useful for
disentangling the relative sign between the scalar ($\kp$) and a
potential pseudoscalar ($\kpt$) top-Higgs coupling.
We have tested the
sensitivity of the various triple product correlations
by considering three types of
observables: asymmetries, angular distributions, and mean values.
Using these observables, we have examined several benchmark
scenarios, focusing in particular on the SM ($\kp=1$ and $\kpt=0$)
and on two ``$\mathrm{CP}$-mixed'' scenarios ($\kp=1$ and $\kpt=\pm1$).
\par

Through the use of spinor techniques we have written the expression
for the differential cross section of the full process in such a
manner that the production and the decay parts are separated, although
connected by the spin vectors of the top and antitop, which are given
in terms of the momenta of the leptons in the final state. Moreover,
we have identified the terms linear in $\kp$ and $\kpt$ as those
involving TPs. Among these, we have explored the three that do not
involve the momenta of the incoming quarks/gluons and at the same time
incorporate both spin vectors: $\epsilon_1\equiv \TPa$,
$\epsilon_2\equiv \TPb$ and $\epsilon_3\equiv \TPc$.\par

We have found
that $\epsilon_{1,2,3}$ allow one to distinguish between the
$\mathrm{CP}$-mixed scenarios by more than $\sim 20\sigma$ in the
case of asymmetries and $\sim 10\sigma$ in the case of mean values
when $1 \times 10^5$ simulated events are used. Furthermore, we have
shown that the angular distributions associated with these TPs are
also sensitive to the values of $\kp$ and $\kpt$, exhibiting a phase
shift that varies according to the values taken by these couplings.
By exploring TPs that incorporate the momenta of the
Higgs and the leptons instead of the spin vectors, we have concluded that
the observables studied here appear to be more sensitive when the spin
vectors are used.\par

We have also proposed a combination of the TPs,
$\epsilon_4\equiv \epsilon_3-\epsilon_2$,
which has a greater sensitivity than $\epsilon_1$-$\epsilon_3$.
With $1\times 10^5$ events, for example, the asymmetry associated
with this TP gives an effective separation 
between the $\mathrm{CP}$-mixed scenarios that
exceeds those coming from $\epsilon_1$-$\epsilon_3$
by at least $2.8\sigma$.  
When a similar
combination is constructed by using the leptons' momenta instead of the
spin vectors ($\epsilon_8$), the sensitivity in the asymmetry is
decreased by $3.1\sigma$ compared to the asymmetry associated
with $\epsilon_4$ for
the same number of events, giving values comparable with those
obtained for the asymmetries of $\epsilon_2$ and $\epsilon_3$. \par

Taking into account the challenge of reconstructing the top and
antitop momenta due to the presence of two neutrinos in the final
state, we have proposed and tested two TP correlations that avoid this
difficulty. The first one is obtained by replacing the $t$ and $\tbar$
momenta by the $b$ and $\bbar$ momenta ($\epsilon_9$), whereas the second
includes the visible part of the $t$ and $\tbar$ momenta
($\epsilon_{10}$). We have found that the latter is the more
sensitive of the two,
leading to a separation between the $\mathrm{CP}$-mixed cases
of $\sim 16\sigma$.\par

Finally, we have discussed the experimental 
feasibility of the most sensitive
observables proposed here. We have found that with $5\times 10^3$ and
$1\times 10^4$ events, respectively, the asymmetries
associated with $\epsilon_4$ and $\epsilon_{10}$ are still
useful for testing the hypotheses $(\kp=1,\kpt=\pm 1)$, giving rise to
separations of order $\sim 6\sigma$. These numbers of events
are within reach in the short term at the LHC,
so that these observables could in principle be used
to test the relative sign of $\kp$ and $\kpt$ within that context.

\acknowledgments
This work has been partially supported by ANPCyT under grants No. PICT
2013-0433 and No. PICT 2013-2266, and by CONICET (NM, AS). The work of
DC, EG and KK was supported by the U.S.  National Science Foundation under Grant
PHY-1215785.
%%%%%%%%%%%%%%%%%%%%%%%%%%%%%%%%%%%%%%%%%%%%%%%%%%%
\setlength{\bibsep}{10pt}
\bibliography{ttbarH_jhep_biblio}
\bibliographystyle{JHEP}
\end{document}